\documentclass[11pt]{article}
\usepackage{amsfonts}

\usepackage{amssymb}
\usepackage{latexsym}
\usepackage{graphicx}
\usepackage[english]{babel}

\begin{document}
\input epsf

\makeatletter
\@addtoreset{equation}{section}
\makeatother


 \begin{center}
{\LARGE Black Holes and Beyond}
\\
\vspace{18mm}
{\bf    Samir D. Mathur
}
\vspace{8mm}

Department of Physics,\\ The Ohio State University,\\ Columbus,
OH 43210, USA\\ 
\vskip 2 mm
mathur@mps.ohio-state.edu\\
\vspace{10mm}

\end{center}

\thispagestyle{empty}

\def\p{\partial}
\def\h{{1\over 2}}
\def\be{\begin{equation}}
\def\bea{\begin{eqnarray}}
\def\ee{\end{equation}}
\def\eea{\end{eqnarray}}
\def\d{\partial}
\def\la{\lambda}
\def\eps{\epsilon}
\def\bb{\bigskip}
\def\mm{\medskip}
\newcommand{\dm}{\begin{displaymath}}
\newcommand{\edm}{\end{displaymath}}
\renewcommand{\b}{\tilde{B}}
\newcommand{\gm}{\Gamma}
\newcommand{\ac}[2]{\ensuremath{\{ #1, #2 \}}}
\renewcommand{\ell}{l}
\newcommand{\z}{\ell}
\newcommand{\newsection}[1]{\section{#1} \setcounter{equation}{0}}
\def\bb{$\bullet$}
\def\Qbar{{\bar Q}_1}
\def\QPbar{{\bar Q}_p}
\def\q{\quad}
\def\bn{B_\circ}
\def\sq{{1\over \sqrt{2}}}
\def\z{|0\rangle}
\def\o{|1\rangle}
\def\sqi{{1\over \sqrt{2}}}

\let\a=\alpha \let\b=\beta \let\g=\gamma \let\d=\delta \let\e=\epsilon
\let\c=\chi \let\th=\theta  \let\k=\kappa
\let\l=\lambda \let\m=\mu \let\n=\nu \let\x=\xi \let\r=\rho
\let\s=\sigma \let\t=\tau
\let\vp=\varphi \let\vep=\varepsilon
\let\w=\omega      \let\G=\Gamma \let\D=\Delta \let\Th=\Theta
                     \let\P=\Pi \let\S=\Sigma

\def\h{{1\over 2}}
\def\t{\tilde}
\def\r{\rightarrow}
\def\nn{\nonumber\\}
\let\bm=\bibitem
\def\Kt{{\tilde K}}
\def\b{\bigskip}

\let\p=\partial
\def\u{\uparrow}
\def\d{\downarrow}

\begin{abstract}

The black hole information paradox forces us into a strange situation: we must  find a way to break the  semiclassical approximation  in   a domain where no quantum gravity effects would normally be expected. Traditional quantizations of gravity do not exhibit any such breakdown, and this forces us into a difficult corner: either we must give up quantum mechanics  or we must accept the existence of troublesome `remnants'.   In string theory, however, the fundamental quanta are extended objects, and it turns out that the bound states of such objects acquire a size that grows with the number of quanta in the bound state. The interior of the black hole gets completely altered to a `fuzzball' structure, and information is able to escape in radiation from the hole. The semiclassical approximation can break at macroscopic scales due to the large entropy of the hole: the measure in the path integral competes with the classical action, instead of giving a subleading correction. 
Putting this picture of black hole microstates together with ideas about entangled states leads to a natural set of conjectures on many long-standing questions in gravity:  the significance of Rindler and de Sitter entropies, the notion of black hole complementarity, and the fate of an observer falling into a black hole.

\end{abstract}
\vskip 1.0 true in

\newpage
\renewcommand{\theequation}{\arabic{section}.\arabic{equation}}

\def\p{\partial}
\def\r{\rightarrow}
\def\h{{1\over 2}}
\def\b{\bigskip}

\def\nn{\nonumber\\ }

\section{The black hole information paradox}
\label{intr}\setcounter{equation}{0}

General relativity and quantum mechanics are two basic pillars of modern theoretical physics. But in 1974 Stephen Hawking argued that these two theories could never be joined together \cite{hawking}. This argument is termed the `black hole information paradox'. 

In outline, the problem can be stated as follows. In any theory of gravity, it is hard to prevent the formation of black holes. Once we have a black hole, an explicit computation shows that the hole slowly radiates energy by a quantum mechanical process. But the details of this process are such that when the hole disappears, the radiation it leaves behind cannot be attributed any quantum state at all. This is a violation of quantum mechanics.

Many years of effort could provide no clear resolution of this problem. The robustness of the paradox stems from the fact that it uses no details of the actual theory of quantum gravity. Thus one of our assumptions about low energy physics must be in error. This, in turn, implies that resolving the paradox should teach us something fundamentally new  about the way that physics works.

Progress over the past several years has suggested a  resolution of the paradox using string theory. And this resolution does come with a basic change in our view of what space-time is,  and when the low energy picture of spacetime becomes invalid. In this article we will strive to give an overview of these developments. We will also discuss some natural conjectures that follow from the new picture of spacetime; these conjectures address several questions about gravity that had been puzzling for many decades. 

\subsection{The semiclassical approximation and its breakdown}\label{infosec}

It is generally accepted that the complete theory of nature must include a quantization of gravity. While quantizing gravity has been a difficult proposition, this fact has not impeded our ability to do everyday physics. The reason for this is our belief in the semiclassical approximation: {\it under appropriate conditions, we can ignore the quantum fluctuations of the gravitational field}. The question then becomes: what are these `appropriate conditions'?

In general relativity we describe the gravitational field by the curvature of spacetime, so we can characterize the strength of this field by a length scale: the curvature radius $R_{curv}$. We can compare this length scale to the natural length scale made out of the fundamental constants $c, \hbar, G$; this gives the planck length $l_p=({\hbar G/c^3})^{1/2}\sim 10^{-33}\, {\rm cm}$. On the surface  of the earth we have $R_{curv}\sim 10^{14}\, {\rm cm}$, so $R_{curv}$ is much larger than $l_p$. The wavelengths $\lambda$ of particles that we study in the lab are also much larger than $l_p$. It is generally believed that whenever
\be
R_{curv}, \lambda ~\gg ~ l_p
\ee
we will be in the semiclassical approximation; i.e., we can ignore the quantization of gravity and represent the gravitational field by just a classical curved spacetime over which we do quantum mechanics. 
Conversely, when $R_{curv}$ or $\lambda$ become comparable to $l_p$, we can expect that the semiclassical approximation will break down, and details of quantum gravity will become important.

In these terms, the Hawking argument says the following:

\b

{\it In our theory of gravity there must be a second mode of breakdown of the semiclassical approximation, which has nothing to do with $R_{curv}$ or $\lambda$ becoming as small as $\sim l_p$. If there is no such second  mode, then one of the following must happen:}

\b

(i) {\it We lose quantum mechanics, i.e., wavefunctions do not evolve unitarily by the Schrodinger equation} $\psi(t)=e^{-iHt}\psi_0$.

\b

(ii) {\it The theory admits `remnants': regions of space where we can hold an unbounded number of quantum states within a finite volume and within a finite energy budget.}

\b

It would be unfortunate if we were pushed to possibility (i), since quantum mechanics has worked perfectly in all other situations. Some relativists have tried to make peace with possibility (ii), but there are serious difficulties with the quantum field theory of remnants: for example it is hard to avoid divergences in loop amplitudes where each available state tends to give a contribution.

 If we wish to admit neither possibility (i) nor possibility (ii), then we must find the second breakdown mode of the semiclassical approximation. This is what had proved so difficult: traditional quantum gravity stubbornly resisted manifesting any nontrivial effects when $R_{curv}, \lambda$ were much larger than $l_p$. And at first the same appeared to be the case with string theory: strings are planck sized objects, and it seems appropriate that we see no evidence of `stringyness' till we reach down to lengths $\sim l_p$.  
 
 But as we will see below, there {\it is} an effect in string theory that will generate the required breakdown of the semiclassical approximation. How is this possible, when dimensional considerations give the natural length scale for quantum gravity effects as $\sim l_p$ ? The situation that Hawking used for his argument -- a large black hole viewed in a `good slicing' -- has $R_{curv}, \lambda\gg l_p$. But a large black hole is made of a large number of quanta $N$, and we can ask if quantum gravity effects are really characterized by the length scale $\sim l_p$ or by some scale $\sim N^\alpha l_p$ for $\alpha>0$. If the latter is the case, then we can break the semiclassical approximation without having high curvatures or high energy quanta;  we can take a large number of {\it low} energy quanta and hope for new effects when the number of quanta is sufficiently large. We will see that the new effects we get can resolve the information paradox. But they are also likely to  be of importance in other situations where we have a large number of quanta, as for example in the early Universe. Thus resolving the paradox is of fundamental importance to theoretical physics.
 
 \subsection{Overview of the article}

 In this article we will do the following:
 
 \b
 
 (a) In section \ref{stwo} we will discuss the Hawking argument in more detail. Many people had concerns about the validity of the argument itself, and so believed that  it need not be taken too seriously. In several  cases this belief could be traced back to one issue: the notion of `small corrections'. Hawking gave a leading order computation of the black hole radiation process, but could it be that subleading corrections, if computed carefully, would alter his conclusion that quantum mechanics was violated? These subleading corrections would be small for each radiated quantum, but there are a large number of radiated quanta, so one might hope for  a nontrivial change in the overall state of the radiated quanta. 
 
 It was recently shown, however, that small corrections do {\it not} alter Hawking's conclusion \cite{cern}. The proof of this fact uses the strong subadditivity property of quantum entanglement entropy,  a somewhat nontrivial result in quantum information theory. Thus it may not be intuitively obvious that the small corrections are indeed irrelevant.  In the end, however, one finds that  the  statement of the information paradox given in section \ref{infosec} can be made fully rigorous. 
 
 \b
 
 (b) In section \ref{sthree} we turn to string theory, where the fundamental particles are {\it extended} objects like strings and branes. The tensions of these objects are order planck scale however, so at low energies a string or brane  is planck size, and does not affect the Hawking argument by itself. But suppose we take $N$ 5-dimensional branes, and bind to them a string. Then the string acquires an effective tension  that is ${1\over N}$ times its normal tension, a phenomenon we term `fractionation'. Here we have a clue that string theory might be different from other theories of gravity: if we make a bound state of  a large number of strings and branes, the resulting low tension objects may stretch far and make the bound state have a size that grows with $N$, instead of staying at a size $\sim l_p$.
 
 Further progress is achieved by making explicit solutions describing particular states of the black hole. Gedanken arguments indicated  that a black hole should have $Exp[S_{bek}]$ states, where the Bekenstein entropy is given by $S_{bek}={A\over 4}$ \cite{bek}. String theory allows us to count these states, but what is more important here is that this counting process also allows us to list these states in an order of increasing `complexity'. Starting with the simplest states allows us to obtain explicit descriptions of actual microstates of the black hole. 
 
Many such constructions have by now been done, and the lesson in each case is the same. The microstate does not have a horizon or a singularity. Thus the entire structure of the state has been altered away from the traditional expectation. For a simple set of microstates the radiation has been computed as well, and found to agree perfectly with the expected rate of Hawking radiation from those special states. But this radiation does not arise by the process that Hawking envisaged, because the structure of the hole is entirely different; this time the evolution is completely unitary.

\b

(c) Finally, in section \ref{sfour} we turn to the potential implications of this new structure of black hole states. What happens if an object falls into such a state? Is there any approximate sense in which the traditional geometry of the hole can be recovered? Looking at the nature of the generic black hole microstate, it appears plausible that a massive infalling object excites collective modes of the microstate. Such modes are quite insensitive to the precise choice of microstate, and an effective universal behavior can emerge for the Green's functions of operators that excite these modes. We will argue that these Green's functions can be interpreted as Green's functions of an effective geometry that agrees with the geometry of the traditional black hole.

If this argument is correct, it suggests a novel view of how effective spacetime geometry arises: the highly complex $Exp[S_{bek}]$ states of a hole radiate their detailed information through low energy processes, but upon high energy impacts they present a universal dynamics that can be interpreted as effective spacetime dynamics. This notion allows us to conjecture solutions to some long standing puzzles. What is the significance of the entropy of Rindler space, where there is no matter whatsoever? Similarly, what is the significance of the entropy of de Sitter space, where the horizon can be moved to different locations by change of coordinate frame? Finally, we can ask what these ideas imply for the early Universe: the Big Bang singularity is very similar to the singularity at the center of a black hole, so understanding the information paradox may help us understand very early time Cosmology.

\section{Hawking's argument}\label{stwo}

In this section we will discuss the Hawking argument in some detail.

\subsection{A first pass}

The core of the argument rests on the process of Hawking radiation: the production of particle pairs from the vacuum in the gravitational field of a black hole. To understand the issues involved, it is simpler to first look at the analogous process with electromagnetic fields -- the Schwinger process.

\b

\begin{figure}[htbp]
\begin{center}
\includegraphics[scale=.55]{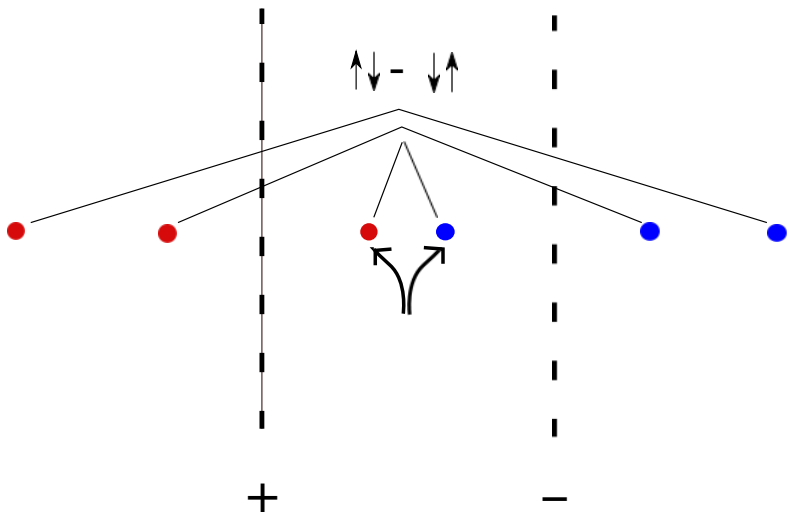}
\caption{{Electron positron pairs are created from the vacuum, and pass through the positive and negative grids. The two members of each pair are entangled with each other, generating an entanglement entropy $S_{ent}=N\ln 2$ between the left and right sides of the figure.}}
\label{f1}
\end{center}
\end{figure}

Consider two wire meshes as shown in fig.\ref{f1}, one positively charged and one negatively charged. The electric field between the meshes leads to a slow process of `particle creation from the vacuum'.  In the vacuum electron-positron pairs are being continually created and destroyed, but the electric field can occasionally separate these two quanta before they re-annihilate, whereupon the electron flies towards (and through) the positive mesh and the positron flies through the negative mesh.

The important thing for us is that the state of the electron and the positron is {\it entangled}. Assume for simplicity that the pairs are created in a total spin zero state
\be
|\psi\rangle_{pair}={1\over \sqrt{2}}\Big ( |\uparrow\rangle_{e^-}|\downarrow\rangle_{e^+}-|\downarrow\rangle_{e^+}|\uparrow\rangle_{e^-}\Big )
\label{one}
\ee
If we divide our space into a left and a right half, then the entanglement entropy between the two sides is
\be
S_{ent}=\ln 2
\ee
The electron and positron drift away from the meshes, and after a while another entangled pair is created between the meshes. 
After $N$ steps of pair creation, we have a set of electrons on the left and a set of positrons on the right. Each electron is entangled with its corresponding positron with the wavefunction (\ref{one}). The total entanglement entropy of the left side with the right side is now
\be
S_{ent}=N\ln 2
\ee
Thus the entanglement between the left and right sides keeps progressively rising.

Hawking found that a similar process of pair creation happens in the {\it gravitational} field of a black hole \cite{hawking,giddingsnelson}. One member of the created pair drifts off to infinity, showing up as the `Hawking radiation' of the black hole. The other member falls into the hole, reducing its mass, so that energy is conserved overall. The two quanta are entangled, so after $N$ steps of pair creation the radiation collecting near infinity is highly entangled with the quanta in the hole. The entanglement at each step is order unity, but its precise value is not relevant; assuming it to be $\ln 2$ as before, we get after $N$ steps an entanglement $S_{ent}=N\ln 2$. 

The Hawking process continues as long as the hole has a horizon that can be described in the semiclassical approximation. This approximation breaks down when the hole reaches planck size. Further evolution will depend on the details of our quantum gravity theory, but Hawking's trap has already been set:

\b

(i) If the hole evaporates away completely, the radiated quanta are in an entangled state, but there is nothing left that they are entangled {\it with}. Thus the state of this radiation cannot be described by {\it any} quantum wavefunction. We could have made the hole by starting with a shell  in a state $|\psi_i\rangle$ that collapsed inside its horizon. A unitary quantum evolution would have given a state $|\psi_f\rangle=e^{-i H t}|\psi_i\rangle$ as the state of the radiation. Thus unitary quantum evolution has to be violated when black holes form and evaporate.

\b

(ii) It is possible that the details of quantum gravity are such that the hole stops evaporating when it reaches planck size, leaving us with a {\it remnant}. This remnant has an entanglement entropy $S_{ent}=N\ln 2$ with the radiation. To be able to have such an entanglement, the remnant must have at least $2^N$ internal states. But we could have started with a collapsing shell of arbitrarily large mass, so $N$ can be made arbitrarily large. Thus remnants would have a curious behavior: they have a bounded radius (order $l_p$) and a bounded energy (order planck mass $m_p$), but they must admit infinitely many internal states.  By contrast, normal field theories only admit a bounded phase space if we limit both the available volume and the available energy. 

\b

Very few people were comfortable with option (i). Many relativists tried to work with option (ii). For example one can imagine that the interior of the remnant hides a baby Universe with an arbitrarily large volume, and the unbounded degeneracy corresponds to states in this baby Universe \cite{baby}.  Alternatively, one may imagine a dynamics where the remnant slowly leaks its energy away by some quantum gravitational process. This leakage has to be slow: since we need to radiate $N$ bits of information with a total energy budget that is only $\sim m_p$, each quantum has energy $\sim {m_p\over N}$, and for large $N$ its wavelength will be very large. In normal field theory, a quantum of wavelength $\lambda$ cannot be radiated by a small object in a time less than $\sim \lambda$, so the overall leakage time of the remnant will be much longer than the Hawking radiation time. Thus the remnants are very long lived objects, and this fact can create many of the same difficulties that would result from completely stable remnants.\footnote{Smolin \cite{smolin} has recently proposed a model of how such leakage could occur if we assume a concept of `relative nonlocality'. Giddings \cite{giddingsmr} has  discussed the possibility of `massive remnants' where some physical process stops the evaporation process before the hole reaches planck scale.} 

String theory aims to be a complete theory of quantum gravity, so option (i) is not acceptable. The theory also seems to have no room for remnants. The gravitational coupling can be smoothly varied between small and large values. At small coupling all states can be explicitly counted, and we find no remnants with unbounded degeneracy. While it is possible in principle that there are remnants are larger coupling, there is no evidence of a sudden transition of this kind.  

This situation led many people to question the Hawking argument itself: perhaps the computation of entanglement has some flaw, so that the entanglement does not in fact monotonically rise with each succeeding emission? After all, when a piece of paper burns away to radiation, the entire evolution is unitary. The entanglement between the paper and the radiation initially goes up: for example the first emitted photon can be in a spin entangled state with the atom it leaves behind. But after about half the paper burns away, the entanglement between the radiated quanta and the remaining paper begins to go {\it down} \cite{page}; for example the atom left behind in the first emission may float out as ash. When the paper is completely gone, there is no entanglement with anything at the original location of the paper; the quanta at infinity are entangled with {\it each other}, thus making a complicated but pure state $|\psi_f\rangle$. 

The case of burning paper indicates that the entanglement of radiation with what is left behind can be a complicated thing to compute, since it involves a large number of quanta and all phases etc. must be carefully tracked. This suggested to many people that it would be easy to find some hole in Hawking's argument, and the quanta radiated from the hole would be in an unentangled state after all. Of course if such a weakness had been found in the argument, the paradox would have disappeared long before now. To understand why Hawking's argument is so robust, let us look at it in more detail.

\subsection{A more detailed look at the Hawking argument}\label{secc}

It is useful to separate the physics of gravity from the purely quantum mechanical aspects of the problem. Thus let us look at Schwinger pair production again, and see what can reduce the entanglement between the left and right sides.

 Let the left side of the Schwinger diagram correspond to the interior of the hole, and the right side to the region near infinity where black hole radiation collects. The space between the meshes is the analogue of the pair-creation region around the horizon, which is a smooth regular piece of spacetime in the traditional black hole geometry. 

After $N$ steps of pair creation, let the overall state of the system be
\be
|\Psi\rangle~=~\sum_{mn}=\t C_{mn}\psi_m\chi_n
\ee
where $\{\psi_m\}$ are a complete basis of states on the left and $\{ \chi_n\}$ are a complete basis of states on the right. We can change basis to make the entanglement diagonal
\be
|\Psi\rangle~=~\sum_i C_i \psi_i \chi_i
\label{five}
\ee
whereupon the entanglement entropy between the left and right sides is given by
\be
S_{ent}=-\sum_i |C_i|^2\ln |C_i|^2
\ee
Since it is this entanglement that creates the information paradox, let us see what could reduce the entanglement.
\b

(a) Suppose we say that we do not know what happens to  stuff that falls deep into the black hole. It could have a complicated evolution, perhaps interacting with the matter shell that initially collapsed to make the hole.  In the Schwinger process this possibility corresponds to a unitary transformation  $\psi_i\r U_{ij}\psi_j$ on the quanta that are outside and to the left of the pair creation region. But any such transformation does not change $S_{ent}$ at all, so we see that the details of what happens inside the black hole do not matter for the computation of entanglement. 

\b

(b) If we cannot remove the entanglement by permuting the stuff inside the hole, then we should see if we can prevent an entangled state of type (\ref{one}) from forming at the step of its production. But this is hard to do since the spacetime around the horizon is gently curved, and there is very little ambiguity in the evolution of quantum fields on such a region.  Similarly, in the Schwinger effect the field between the plates is well known, and if do have this field $\vec E$  then it is hard to make a significant alteration in the state of the created pair.  

 There is always a very small but nonzero  probability that the entire region between the plates undergoes a nonperturbative quantum fluctuation that completely changes its character; in this case of course the pair production at that moment could be very different from (\ref{one}). We mention this possibility because people have tried to argue that once in a while the entire black hole can undergo a huge quantum fluctuation that destroys the semiclassical approximation. And if we lose the semiclassical approximation, then perhaps Hawking's argument would be invalidated.

But this argument has little merit. Even if we destroy the entanglement in one pair produced in the Schwinger process, the entanglement created by the remaining pairs keeps $S_{ent}$ of the same order as before. We would have to destroy the entanglement of virtually every created pair to 
invalidate the Hawking argument, and this is equivalent to saying that the semiclassical black hole geometry is completely wrong. In short,  it does not help to say that the geometry deviates from the semiclassical one `occasionally'. We expect that such  fluctuations  can only occur with a probability that is exponentially small in the mass of the hole, and the corresponding change in entanglement will not affect Hawking's argument.

 \b
 
(c) One way in which we {\it can} change the entanglement created in the Schwinger process is the following. Let us replace the positively charged mesh with a positively charged impenetrable {\it plate}. Now the created electrons do not  escape; instead they collect in the vicinity of the positively charged plate. After a sufficient number of electrons collect here, we can no longer think of the region of pair production as the vacuum. The created pairs no longer have to be in the state (\ref{one}), and the entanglement between the left and right sides need not keep rising. 

Some people thought the black hole would be analogous to this case. If we picture the hole as a ball of radius $R$, then it would seem that  the initial matter shell that made the hole and the infalling members of the created pairs would all have to `fit' in this ball. After enough quanta collect in the hole, would it not be true that the later quanta would be significantly influenced by the earlier ones, invalidating the expression (\ref{one})? 

The answer is: the interior of the black hole is {\it not} like a finite volume ball. The quanta that fall into the hole {\it do} get flushed away just as in the Schwinger model with meshes on each side. This is in  fact the crucial property that makes a black hole different from a star, and it arises because the black hole has a {\it horizon}. To understand this notion, consider the black hole metric
\be
ds^2=-(1-{2M\over r})dt^{2}+{dr^2\over 1-{2M\over r}}+r^{2}{d\Omega_2^{2}}
\label{two}
\ee

\begin{figure}[htbp]
\begin{center}
\includegraphics[scale=.55]{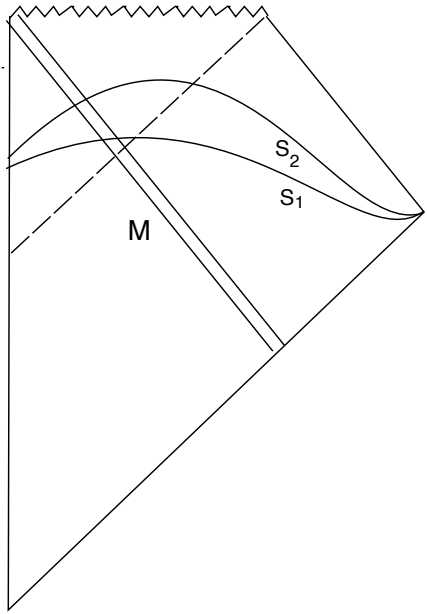}
\caption{{The Penrose diagram of a black hole formed by collapse of the `infalling matter'. The spacelike slices shown give a foliation of the geometry by `good slices'.}}
\label{fthree}
\end{center}
\end{figure}

\begin{figure}[htbp]
\begin{center}
\includegraphics[scale=.18]{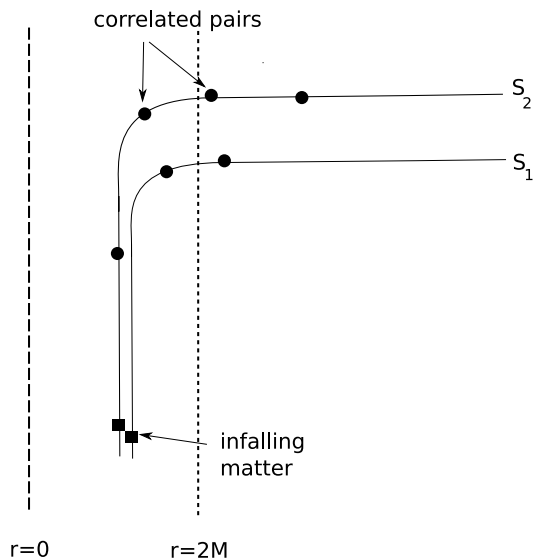}
\caption{{Eddington-Finkelstein coordinates for the Schwarzschild hole. Spacelike slices are $t=const$ outside the horizon and $r=const$ inside.  Curvature length scale is $\sim 3 ~km$ all over the region of evolution covered by the slices $S_i$.}}
\label{ftwo}
\end{center}
\end{figure}

We see that for $r<2M$ the directions $t, r$ interchange roles: the $t$ direction becomes spacelike and the $r$ direction becomes timelike. Thus fact has a crucial impact on the nature of spacelike slices that we must draw to study evolution in the black hole geometry. Such spacelike slices are shown in the Penrose diagram of fig.\ref{fthree}. Since the Penrose diagram scales different regions of spacetime differently, it is somewhat clearer to see these slices in Eddington-Finkelstein coordinates, used in fig.\ref{ftwo}. The horizontal coordinate plots $r$, defined through the requirement  that the area spanned by the angular variables $\theta, \phi$ is $4\pi r^2$.  The vertical coordinate plots 
\be
u=t+r^*=t+r+2M\ln |{r\over 2M}-1|
\ee
The Schwarzschild metric (\ref{two}) becomes
\be
ds^2=-(1-{2M\over r})dudr + 2dudr+r^2 d\Omega_2^2
\label{three}
\ee
which is regular at the horizon. 

We can make a spacelike slice as follows:

\b

(i) For $r>4M$, take the slice as $t=t_1=constant$.

(ii) Inside the hole, let the slice  asymptote to $r=M$. A constant $r$ segment is {\it spacelike} since $r, t$ have interchanged roles inside the horizon $r=2M$.

(iii) Join the above  two parts with a smooth spacelike surface. We can choose this surface so that the proper length $ds$ along the surface agrees with the change $du$ 
\be
{du\over ds}=1 ~\Rightarrow~s=r+M\ln(r-M)
\label{four}
\ee
To understand this condition, suppose we have an infalling scalar field wavefunction at infinity $\phi=Exp[-ik(r+t)]\approx Exp[-iku]$. When this 
wave is seen on our spacelike slice, its wavefronts are separated by distances $\Delta s=\Delta u = 2\pi/k$. Thus if the wave had wavelength $\lambda=2\pi/k\gg l_p$ at infinity, then as observed on our spacelike slice it has the same wavelength $\lambda_{slice}=\lambda\gg l_p$. Recall that one of the conditions for the semiclassical approxomation was that the wavelength of all quanta should satisfy $\lambda\gg l_p$, and we see that all low energy matter infalling into the hole will satisfy this condition on our slice. The other condition for the semiclassical approximation was $R_{curv}\gg l_p$, and from the regularity of the metric (\ref{three}) at the horizon we observe that $R_{curv}\sim M\gg l_p$.

To analyze the evolution we have to now make a `later' slice (fig.\ref{ftwo}):

\b

(i) The part at $r>4M$ is advanced to $t=t_2>t_1$, with no change in its intrinsic geometry.

(ii) `Forward evolution' for the part $r\approx M$ would mean moving towards smaller $r$, since $r$ is the `time' direction inside the hole. We do not want to go to small $r$ since there is a curvature singularity at $r=0$, and we cannot make any reliable statements about the evolution if we get into a region of high curvature. Thus we keep this part of the slice at $r\approx M$.

(iii) We keep unchanged the intrinsic geometry (\ref{four}) of the `connector segment'. From fig.\ref{ftwo} we see that the part of the slice at  $r\approx M$ has to {\it stretch} in this process of evolution. Thus the successive slices in our foliation do not have the same intrinsic geometry; in fact it is crucial that there is {\it no} slicing of the black hole geometry in which the slices all have the same intrinsic structure.  

\b

This stretching leads to the creation of particle pairs: on any given slice we might find a state that we call the `vacuum', but if we deform the slice then we would get a different `vacuum', and the difference is radiated away as real quanta that have been `pair produced from the vacuum'. Looking at the slices we have, we can describe the creation process as follows. The 
left and right parts of the slice keep their intrinsic structure, but the middle gets `stretched'. This creates a pair of entangled quanta. The process repeats, so that the quanta of the previous steps get {\it separated} away, and a new pair is created by the stretching in the middle of the slice. If we look at the entangled pairs on the entire slice, we get a picture analogous to fig.\ref{f1}, where the created quanta are `flushed away' from the creation region, instead of `piling up' there and affecting future pair production steps. Thus we cannot get away from the Hawking problem by regarding the black hole as a normal radiating `ball'.

\b

(d) If we cannot make a significant change to each produced pair, could we manage with {\it small} corrections to {\it each} pair? It is certainly possible that there are small quantum gravity effects, possibly nonlocal,  that make small changes in the  evolution near the horizon; the regularity of the geometry there only says that the {\it leading} order evolution is given by standard `quantum fields on curved space'. The number of created pairs is $\sim (M/m_p)^2\gg 1$. Is it possible that the smallness of the correction at each step can be offset by the large number of pairs, so that we can produce a state of radiation at the end that is not entangled with the planck sized hole near the end of the radiation process?

As we will see below, this issue caused a vast amount of confusion in the literature of the information paradox. The answer is that small corrections {\it cannot} remove the entanglement. But proving this takes some effort, so the result may not be intuitively obvious. A detailed analysis is given in \cite{cern}; here we outline the steps. To set up the problem, consider a given step in the evolution, and write the state on a complete slice through the black hole geometry in a manner analogous to the Schwinger process state (\ref{five})
\be
|\Psi\rangle~=~\sum_i C_i \psi_i \chi_i
\label{fiveq}
\ee
where $\{ \psi_i\}$ are states describing the shell which made the hole as well as all the infalling members of the created pairs, and $\{ \chi_i\}$ are states of the quanta that have already been radiated. Consider the evolution to the next step where one more pair is created. In the leading order Hawking process the state of the quanta that have already been emitted does not change
\be
\chi_i~\r~ \chi_i
\ee
The state $\psi_i$ in the hole evolves as
\be
\psi_i~\r~\psi_i\otimes S^{(1)}
\ee
where $S^{(1)}={1\over \sqrt{2}}\Big ( |\uparrow\rangle_{e^-}|\downarrow\rangle_{e^+}-|\uparrow\rangle_{e^+}|\downarrow\rangle_{e^-}\Big )$ is the state of the created pair. To allow for small corrections, we can add a small admixture of the other basis state 
\be
S^{(2)}={1\over \sqrt{2}}\Big ( |\uparrow\rangle_{e^-}|\downarrow\rangle_{e^+}+|\uparrow\rangle_{e^+}|\downarrow\rangle_{e^-}\Big )
\ee
We will also allow for an arbitrary change of the stuff that is inside the hole; this does not happen in evolution in the standard geometry of the hole (see fig.\ref{ftwo}), but as we saw in point (a) above, such changes should not affect the entanglement computation.  So we lose nothing by allowing the more general evolution
\be
\psi_i~\r ~\psi_i^{(1)}\otimes S^{(1)}~+~\psi_i^{(2)}\otimes S^{(2)}, ~~~~~~~~|\psi^{(1)}|^2+|\psi^{(2)}|^2=1
\ee
Thus the evolution through one step has the form
\be
\sum_i C_i \psi_i\chi_i~\r~\sum_i C_i [\psi_i^{(1)}\otimes S^{(1)}~+~\psi_i^{(2)}\otimes S^{(2)}]\chi_i~\equiv~ \Lambda^{(1)}  S^{(1)}~+~ \Lambda^{(2)}S^{(2)}
\ee
Here
\be
 \Lambda^{(1)}~=~\sum_i C_i \psi_i^{(1)} \chi_i, ~~~ \Lambda^{(2)}~=~\sum_i C_i \psi_i^{(2)} \chi_i, ~~~~~~~|\Lambda^{(1)}|^2+|\Lambda^{(2)}|^2=1
 \ee
If evolution at the horizon has to be be `close' to the leading order semiclassical evolution expected on gently curved spacetime, then we should have only a small admixture of $S^{(2)}$ in the state of the created pair. This is quantified by requiring
\be
|\Lambda^{(2)}|~<~\epsilon, ~~~~~~~\epsilon\ll 1
\ee
Note that the admixture of $S^{(2)}$ at each step can depend in an arbitrary way on the matter that made the hole and the earlier quanta that fell in the hole; all we require is that this admixture be small so that we get the physics of a regular horizon to leading order. It was shown in \cite{cern} that after $N$ steps the reduction in entanglement $\delta S_{ent}$ from the leading order value $S_{ent}=N\ln 2$ is bounded as
\be
{\delta S_{ent}\over S_{ent}}~<~ 2\epsilon
\label{seven}
\ee
Thus we see that small corrections to the evolution of each pair will not help resolve the information paradox.

\subsection{The no-hair `theorem'}

If we cannot escape the Hawking argument with small corrections to the evolution at the horizon, why don't we look for corrections to the black hole geometry which would make order {\it unity} corrections to the state of the created pair? If we can alter the black hole geometry away from (\ref{two}), then the Hawking argument can be avoided. The state of the created pair can now depend on the details of the metric around the horizon, so the state may not be entangled in the manner (\ref{one}). 

\begin{figure}[htbt]
\hskip .5 in\includegraphics[scale=.45]{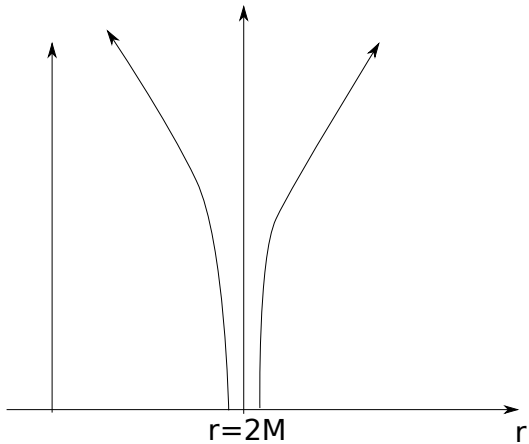}
%
%
\caption{The instability of `outgoing geodesics' at  the horizon. A null geodesic at $r=2M$ headed radially outwards stays stuck at $r=2M$, one slightly outside escapes to infinity while one slightly inside falls to $r=0$.}
\label{fa14}       
\end{figure}

People tried to look for deformations of the black hole geometry -- adding `hair' to the hole -- but so such deformations were found. The essential reason for this failure can be understood by examining the physical nature of the horizon. A massless quantum flying radially outwards at $r=2M$ stays at $r=2M$. If it started at $2M+\epsilon$, then it eventually escapes to $r=\infty$. If it started at $r=2M-\epsilon$, it eventually falls into the singularity at $r=0$. Thus the horizon is an `unstable' place, as depicted in fig.\ref{fa14}. This separation of neighboring geodesics leads to an expansion of the region around the horizon similar to the expansion of de-Sitter space. 

Consider a spacelike slice of the kind we made above, and assume that we have some quanta in the region around the horizon. As we evolve to later slices, the quanta flow off to infinity or fall in towards $r=0$. The region around the horizon becomes the vacuum, and further evolution creates the usual entangled pairs assumed in the leading order Hawking process. Since the black hole metric quickly settles back to the vacuum solution (\ref{two}), we say that `black holes have no hair'; i.e., we have no deformations of (\ref{two}) that persist for times comparable to the duration of the evaporation process \cite{nohair}.

The continual stretching of geodesics away from the horizon removes any distortions from the vicinity of the horizon, but we can still be uncomfortable with the notion that the state left at the horizon after this stretching is the {\it vacuum}. After a small part of the evaporation process has occurred, we find that the created pairs are arising from modes that were {\it transplanckian} before the hole formed; i.e., they had wavelengths $\lambda\ll l_p$ before the hole formed and they have been stretched to $\lambda\sim M$ at the point where they give rise to a created pair.  Could it be that when we pull transplanckian modes from the vacuum by stretching, we obtain not the vacuum but some corruption of the vacuum, and the standard Hawking pair creation is correspondingly altered?

But we cannot get away from the Hawking argument so easily.  In inflation, a marble sized Universe expands to a size over 3000 Mpc; thus atomic wavelengths today started off as transplanckian before inflation. But atomic scale wavemodes show no novel physics that deviates from the traditional expected physics for those modes. Further, as we will see now, it is possible to formulate the Hawking argument that makes no reference to transplanckian modes at all; the physics of such modes is bypassed by using the `no-hair theorem'. 

\subsection{The Hawking argument summarized}

Let us put together all that we have seen above into the following argument. Consider the slab of spacetime between two `nice' spacelike slices separated by $\Delta t\sim M$. In this evolution the leading order Hawking computation yields a new  entangled pair of the form (\ref{one}).
Such entanglement will force us to  either information loss or remnants. To avoid this conclusion, we have three choices:

\b

(a) We can somehow change the state on the initial slice, so that it is not close to the vacuum.

(b) We can change the evolution between the slices, so that it is not the one given by the normal `quantum fields on curved space' evolution.

(c) We change neither the initial state nor the evolution to leading order, but somehow try to encode delicate correlations in small corrections to the created pairs, hoping that this will remove the entanglement.

\b

But as we have seen above, each of these escape routes has faced problems:

\b

(a') When people tried to deform the black hole geometry, they did not succeed; this led to the belief that `black holes have no hair'. If we keep the black hole metric but try to add some quanta to the geometry, the persistent stretching at the horizon rapidly moves these quanta away, leading us back to the vacuum state at the horizon.

(b') Since $R_{curv}, \lambda\gg l_p$ at the horizon, we are in the semiclassical approximation, so it is not clear {\it why} the evolution should be altered. If someone postulated a different evolution, he would have to explain why the evolution should not be altered in other places where we use the semiclassical approximation, for example in a lab on earth or on a slice through the galaxy.

(c') Here we have the inequality (\ref{seven}), which shows that small corrections {\it cannot} make any significant reduction in the entanglement.

\b

Given these difficulties, many relativists had settled on some version of the remnant scenario, despite its problems. But as we will see, string theory led to a radical change in our understanding of black holes and the structure of spacetime. We will turn to these developments next.

\section{The structure of black holes in string theory}\label{sthree}
 
 String theory begins with the postulate that the fundamental quantum is not a point particle but a string loop. But at the low energies appropriate for the semiclassical approximation the loop is very tiny -- planck size -- so it is unclear how this structure can have any influence on the Hawking argument. Indeed,  the 3-point function of three strings at low energy matches the 3-graviton vertex of general relativity, as it should if string theory is to reproduce low energy gravity. But as we will see now, there {\it is} a way that string theory differs from other theories of gravity, and the difference comes when we consider the bound state of a {\it large} number of quanta. The fundamental objects in the theory exhibit an effect called fractionation, which generates very low tension `effective objects' when we make bound states of many quanta.
 
 \subsection{Fractionation}
 
 To make a black hole in string theory we have to take the fundamental objects in the theory, and bind them together to make a massive source for the gravitational field. Since the theory lives in 10 dimensions, we imagine that 6 directions are compactified on small circles. We can take a string and wrap it $n_w$ times around a compact circle. This generates a `point mass' from the viewpoint of the noncompact dimensions; this mass also carries a charge -- the `winding charge $n_w$. Alternatively, we can take massless gravitons and let them circulate  as `momentum modes' around the compact direction. If we take a graviton with $n_p$ units of momentum, then from the viewpoint of the noncompact directions we get  a pointlike mass  with `momentum charge'. 
 
 It is also easy to make bound states of these charges: if we bind a momentum mode to a winding string then we simply get a string carrying the momentum as travelling waves. What will be of interest to us is the {\it energy gap} for excitations in our system.
  
 \subsubsection{Energy levels on the `multiwound' string'}
 
   \begin{figure}[htbt]
\hskip .5 in\includegraphics[scale=.45]{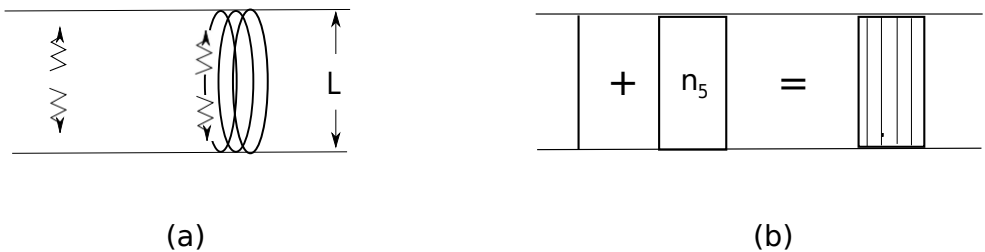}
%
%
\caption{(a) Momentum modes by themselves have energies $\sim {1\over L}$, but when bound to a multiwound string with winding $n_w$ the minimum allowed energy drops by a factor $n_w$.  (b) Duality maps this simple observation to an interesting fact about branes: D1 branes bound to many D5 branes give rise to `fractional tension' branes. }
\label{fa3}       
\end{figure}

 Let the compact direction have length $L$. The minimum energy of a momentum mode is ${2\pi\over L}$. If we wish to make an excitation with no net charge we can take a momentum mode and an `anti-momentum mode' -- i.e. a graviton going the other way around the circle. The total energy would be 
 \be
 \Delta E={2\pi\over L}+{2\pi\over L}={4\pi\over L}
 \ee
 But now consider binding these excitations to a string which has winding number $n_w$. The total length of this string is $L_T=L n_w$, so its vibration energies are in multiples of ${2\pi \over n_w L}$. The minimum energy excitation with no net momentum charge again consists of one left and one right moving vibration, with total energy
 \be
 \Delta E={2\pi\over n_w L}+{2\pi\over n_w L}={4\pi\over n_w L}
 \label{ten}
 \ee
 Interestingly, the energy of a momentum-antimomentum pair has dropped by a factor $n_w$ when it is bound to $n_w$ units of string winding charge, as compared to the case where it is not bound to anything. If we have $n_p$ units of momentum bound to the string, then we should write the total momentum as
 \be
 P={2\pi n_p\over L}={2\pi n_wn_p\over L_T}
 \label{nine}
 \ee
 so that there are effectively $n_wn_p$ `fractional' excitations on a `long' string of length $L_T$ \cite{dasmathur}.

 \subsubsection{Fractional tension branes}

 The above may look like a trivial point, since it is true for the behavior of any string-like object in any quantum theory. But the interesting point about string theory is the presence of {\it duality symmetries} that map charges in the theory to other charges. We can make such a map so that 
 \bea
 n_w ~{\rm units ~of ~string ~winding} ~&\r& ~ n_w ~ {\rm D5~branes}\nn
  n_p ~{\rm units ~of ~momentum} ~&\r& ~ n_p ~ {\rm D1~branes}
  \label{qwone}
 \eea
While a D1 brane by itself has planck scale tension, the above map suggests that when it is bound to $n_5$ D5 branes, we will effectively have a D1 brane with fractional tension: a  tension equal to ${1\over n_5}$ of its normal value. If we bind $n_1$ D1 branes to $n_5$ D5 branes, then  
 (\ref{nine}) suggests that we should think of the bound state as being composed of $n_1 n_5$ strands of an `effective string' with fractional tension. 
 
 If the length of the wrapping cycle is  $L$, and if we join all the strands of this effective string into one `multiwound' strand, then we get a string with effective length $L_T=n_1n_5 L$. If we excite this effective string, then we will get  momentum-antimomentum excitations which can have energy as low as \cite{maldasuss}
 \be
 \Delta E={4\pi\over n_1n_5 L}
 \label{el}
 \ee
 Note that when  two kinds of charges were involved (winding and  momentum) a single charge appeared in the denominator (eq.(\ref{ten})), but when three charges were involved (D1, D5 momentum) then we get {\it two} charges in the denominator (eq.(\ref{el})). We call these cases `single fractionation' and `double fractionation'. The general moral is: making bound states of appropriate kinds of charges allows us to make excitations that are very light compared to what they would be if they were {\it not} part of the bound state. Before we see the consequences of this fact, let us pause a little to recall the story of black hole entropy.

 \subsubsection{The entropy of black holes}
 
 Bound states of branes have been very important in the story of black holes in string theory.  Black holes have an entropy $S={A/4G}$, but what is this entropy counting? The first results about black holes from string theory were directed at answering this question. We can look at supersymmetric states, whose degeneracy does not change with coupling. We take a supersymmetric bound state  at weak coupling, and count its degeneracy. We can then imagine that the coupling is increased till the bound state curves space enough to make a black hole. If the horizon area $A$ of this hole gives a Bekenstein entropy that agrees with the weak coupling count of states, then string theory would have passed a crucial test as a theory of quantum gravity. 
 
We can use the notation developed in the above section to recap the work on black hole entropy.  Consider the bound state of string winding plus momentum.   The momentum (\ref{nine}) was carried on the string as $n_wn_p$ units of vibration, and the number of ways of partitioning this momentum among different harmonics of vibration gives an entropy \cite{sen}
 \be
 S_{micro}=4\pi\sqrt{n_wn_p}
 \label{fourt}
 \ee
 If we make a black hole carrying the mass and charges appropriate to $n_w$ units of string winding and $n_p$ units of momentum, then we can deduce its Bekenstein-Wald entropy from its horizon geometry, finding \cite{dabholkar}
 \be
 S_{bek}=4\pi\sqrt{n_wn_p}=S_{micro}
 \label{sixt}
 \ee
 This computation shows the remarkable consistency of string theory: if we allowed the string to vibrate in 9 dimensions instead of 10, then we would not get the agreement (\ref{sixt}). Thus correctly quantizing the free string (which forces us to be in 10 dimensions) helps us reproduce the correct black hole entropy when we consider a complicated bound state in the theory.

\begin{figure}[htbt]
\hskip .5 in\includegraphics[scale=.45]{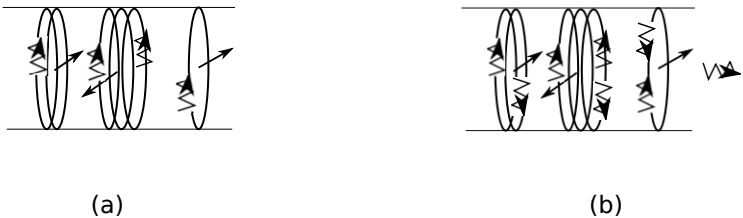}
%
%
\caption{(a) The CFT states of the extremal D1-D5-momentum hole are given by letting the effective string break into loops  in all possible ways, and  distributing the momentum in arbitrary ways over these loops. (b) Allowing the momentum carrying vibrations to run  in both  direction along  the effective string gives states of  the non-extremal hole; collisions of oppositely moving vibrations leads to emission that has a rate that exactly matches the Hawking emission rate from the corresponding black hole.}
\label{fa5}       
\end{figure}

 Similarly, consider the bound state of $n_1$ D1 branes and $n_5$ D5 branes. We had seen that this was described by an `effective string' with winding $n_1n_5$. If we add $n_p$ units of momentum as vibrations to the effective string, then the entropy of vibrations gives (fig.\ref{fa5}(a))
 \be
 S_{micro}=2\pi\sqrt{n_1n_5n_p}
 \ee
 A black hole with the same charges gives \cite{sv}
 \be
 S_{bek}=2\pi\sqrt{n_1n_5n_p}=S_{micro}
 \label{fift}
 \ee
 so we again get an exact agreement. Eq. (\ref{fift}) was in fact the first exact agreement on entropies that was obtained. A precursor to such computations was \cite{larsen} where the entropy of 4-charge microstates was computed by partitioning momentum among harmonics; 
 in this case the result differed by  factor of $2$ from the Bekenstein entropy since the quantization of monopole charge was taken to be the standard one in field theory rather than the one obeyed by D-branes in string theory. A little earlier it had been argued that the Bekenstein entropy could be reproduced by the excitations of a string if we assumed that the tension of the string was appropriately renormalized in the gravitational field of the extremal hole \cite{cvetic}. 
  
The above bound states give `extremal' holes since the mass of the state equals the charge; in this situation the temperature is zero and the hole does  not radiate. To get radiation,  we can add both momentum and anti-momentum modes to the D1D5 effective string.  The microscopic state count again reproduces the Bekenstein entropy \cite{hms}. Since the mass is now {\it more} than the charge,  the extra energy can come off the effective string in the form of  radiated gravitons when momentum and anti-momentum modes collide (fig.\ref{fa5}(b)). This rate of such radiation from the microstate  is found to agree exactly with the spectrum of low Hawking radiation from the black hole with the same mass and charges \cite{comparing}
\be
\Gamma_{micro}~=~\Gamma_{hawking}
\label{tw}
\ee

\subsubsection{The size of bound states}

While the above agreements of entropy and radiation rates are deeply interesting, they do not by themselves resolve the information paradox. The LHS of (\ref{tw}) is computed by a unitary process involving vibrations of an effective string:  the radiation is emitted just like the radiation from a piece of burning paper, and there is no information problem.  The RHS of (\ref{tw})  is computed by  Hawking's process that leads to an ever growing entanglement between the radiation and the hole, and this process {\it does} lead to the information problem at the end of the evaporation process.

To address the information question we need to see if the structure of the black hole in string theory could be different from the traditional one.   Here we get a hint from the fractionation expressions (\ref{ten}),(\ref{el}).
To make a large black hole we have to make a bound state of large  numbers of charges. But binding many charges together generates  a large degree of `fractionation', so that the fundamental objects of string theory come with not planck scale tension but much lower effective tensions. These low tension objects may be able to stretch far, giving the entire bound state a significant effective size. An estimate of this size $D$ was made for the D1-D5-momentum bound state in \cite{emission}, giving
\be
D~\sim~ \Big ({g^2 \alpha'^3 \sqrt{n_1n_5n_p}\over RV}\Big )^{1\over 3}
\label{thir}
\ee
Here $g$ is the string coupling and $\alpha'$ is defined through the string tension $T=1/(2\pi \alpha')$. The D1 branes wrap a circle of radius $R$ and the D5 branes wrap a 4-torus of volume $V$ times the circle of radius $R$.  But the scale (\ref{thir}) is exactly of the order of the horizon radius $R_s$ of the D1-D5-momentum black hole
\be
R_s= \Big ({g^2 \alpha'^3 \sqrt{n_1n_5n_p}\over RV}\Big )^{1\over 3}
\ee
This suggests that something remarkable happens in string theory. The effective size of brane bound states {\it grows} with the number of charges in the bound state, in such a way that this size is always of order the horizon size of a black hole with the same mass and charges. If such is indeed the case, then the structure of the hole will be completely altered from the traditional assumption (\ref{two}), and we need not have the progressive entanglement of radiated pairs that creates the Hawking paradox.

To investigate this in more detail, we should look at the structure of brane bound states in more detail. But before we do that, we will take a look at the idea of AdS/CFT duality, which emerged soon after this stage in our understanding of black holes. This duality made very deep changes to our view of spacetime, but it also led to several popular confusions about how string theory would resolve the information paradox.

\subsection{AdS/CFT duality}

 Consider a bound state of many D-branes. The gravitational field of these branes will deform the space near the branes. If we take the D1-D5 bound state considered above, or just a collection of D3 branes, the near region geometry is a piece of anti-de Sitter spacetime (AdS). The gravitational field very near the branes would be intense, and we would expect the usual difficulties with understanding quantum gravity at strong coupling
 at the location of the branes. Gravitons and other closed string modes are described by tubes that have their ends on the branes, and we would need to study the detailed interactions between all these closed string modes to understand the physics near the branes.

 The remarkable thing about AdS/CFT duality is that we obtain an explicit description of this strong gravitational physics in a way that involves {\it no} gravity. The interaction between D-branes is given by open strings that have their ends on the branes. A loop of open strings looks like a {\it closed} string. Thus we can describe the mess of closed string gravitational interactions by a set of open string diagrams. The complexity of the interactions has not decreased, but there is a natural conjecture for the physics that leads to the open string diagrams: it is a conformally invariant {\it field theory}. For the case of D3 branes, the field theory is ${\cal N}=4$ supersymmetric Yang-Mills theory, while for the D1D5 bound state we have an `orbifold field theory' \cite{maldacena}.

  \begin{figure}[htbt]
\hskip .5 in\includegraphics[scale=.35]{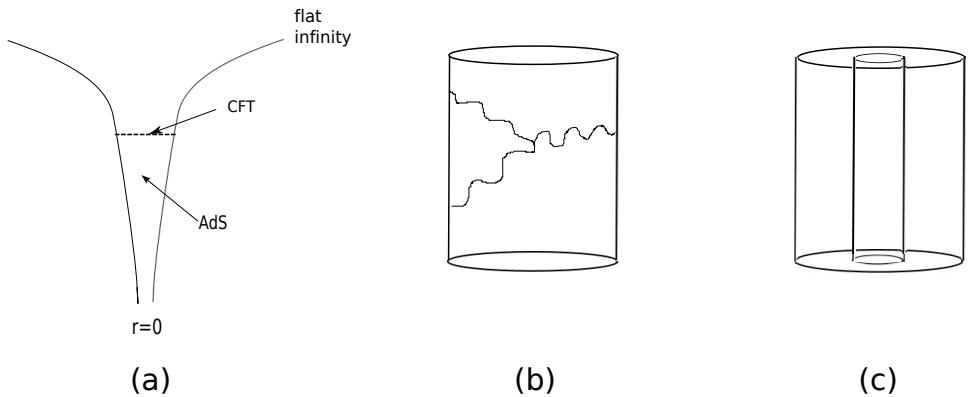}
%
%
\caption{(a) Branes create a geometry that is AdS in the `near' region; the dual CFT lives on a boundary placed anywhere in the AdS region, and describes gravity in all the region below it. (b) The singularity at $r=0$ can be avoided by moving to global AdS, where a 3-point function is computed by a simple path integral with no singularities. (c)  If we have enough energy in global AdS, we make a black hole, and then we face the difficulties of Hawking's argument again. (The vertical direction is time, and the surface of the inner cylinder is the black hole horizon.)}
\label{fa1}       
\end{figure}

 How should we use this field theoretic description of gravitational physics? In fig \ref{fa1}(a) we depict the geometry created by the branes, with the near-brane region being given by a piece of AdS spacetime. We draw an imaginary surface at some point in the AdS region and consider our field theory placed on that surface. This field theory is supposed to capture the gravitational physics of the entire region below the surface.
 
 Using the duality in this form is a little confusing, since on the gravity side the geometry looks singular at $r=0$ where the branes were originally placed. Instead of taking a piece of AdS we can take {\it global} AdS, which is a homogenous space with no singularity (fig.\ref{fa1}(b)). The CFT is taken to live on the boundary of this regular space, and gives a dual description of all the gravitational physics on the space. We can take Euclidean or Lorentzian signature for AdS, and the field theory then has the same signature.
 
The duality in this form has been extremely useful, and has led to many insights into field theory and gravity. But for our black hole problem, there are some  puzzling points that need clarification:
 
 \b
 
 (i) The D3 brane bound state is unique (more precisely, it is a single supermultiplet of 256 states). But the D1D5 brane bound state has an entropy $S=4\pi\sqrt{n_1n_5}$ (this follows by duality from (\ref{fourt})).  Thus there are $Exp[S]$ different bound states with the same mass and charges. Which of these are represented by the AdS geometry? If there are $Exp[S]$ different dual gravity solutions, where should we look for the difference between the solutions? Near $r=0$, or elsewhere in the AdS region? 
 
 (ii) With global AdS the duality works perfectly for low energy correlators like 3 and 4 point functions of gravitons. But suppose we take the case of Lorentzian signature and put enough energy in the AdS to make a black hole (fig.\ref{fa1}(c)). The black hole has a temperature and an entropy that we may identify with the temperature and entropy of the state in the dual field theory. The CFT is perfectly unitary, but for the black hole people just wrote down the metric for the AdS-Schwarzschild hole 
 \be
 ds^2=(r^2+1-{M\over r^2})dt^2+{dr^2\over r^1+1-{M\over r^2}}+r^2d\Omega_3^2
 \ee
 Applying the Hawking argument to this metric, we have the same information loss problem as before.  How can the duality work?
 
 \b

 It is important to consider these questions, since in the early days of AdS/CFT they were often {\it bypassed}. Ignoring this issue led to some arguments that are actually fallacious:
 
 \b
 
 (a) It was argued by some that the CFT describes the correct microscopic physics of black holes, and thus it is the field theory degrees of freedom whose count  should reflect the entropy of the hole. The gravity solution is an effective description of all these states, and it makes no sense to look for different `microstates' that distinguish $Exp[S]$ gravitational solutions; i.e. we should get only {\it one} gravitational solution describing {\it all} the microstates, so the black hole will still have no hair.
 
 (b) Others argued that there should be $Exp[S]$ different gravitational microstates for the hole, but there is no reason to go and look for them. Since the gravity theory is dual to a CFT, and the CFT is manifestly unitary, the information paradox must automatically solve itself somehow.  The differences between the gravitational microstates would be so small in any case that it can only be probed by complicated planck scale measurements, and so such differences would have no impact on the low energy physics that we wish to do.  
  
 \b
 
 Neither of these arguments is valid, however, as we now discuss:
  
 \b
 
 (a')  This line of reasoning arose because people could not find the `hair'  which would distinguish different microstates on the gravity side. But the nature of AdS/CFT duality is such that we expect an exact match of quantum states between the two sides; thus if the duality is correct then there {\it must} be different gravitational solutions corresponding to the $Exp[S]$ microstates. Most importantly, if we do not find any `hair' to modify the geometry at the horizon, we will be forced to information loss/remnants by Hawking's argument, and this is a problem since string theory seems to be unable to accommodate either of these options.
 
(b') This argument for resolving the information paradox is fallacious for the following reason. The duality has verified for low energy processes like 3 and 4 point functions, and we have then extrapolated it to the black hole. But if we cannot find an actual  way around the Hawking argument then all we would have learnt is that this {\it extrapolation} is false; string theory fails at the same place where all other quantum gravity theories fail -- at the point where a black hole is formed.

Some people tried to use argument (b) another way: let us {\it define} the gravity theory to be the dual of the CFT. Surely black hole evaporation would have to be unitary in this case? But now we have the opposite problem: the gravitational theory defined by the CFT reproduces low energy graviton correlators, but how do we know that when the energy is increased the theory will really contain a {\it black hole}? A black hole is a very special object: it forms in a time scale $\sim M$, but decays very slowly, over a time $M(M/m_p)^2$. A generic CFT need not have states which are metastable for such long times. In fact there is only one case of AdS/CFT type duality that is explicitly solved: the 1-d matrix models whose dual is approximated at low energies by 1+1 dimensional gravity. The matrix model reproduces low energy gravity correlators perfectly, but while 1+1 gravity theories typically have black holes, the matrix model does {\it not}. The corrections that distinguish the matrix model from low energy gravity are small for low energy processes, but become order unity  when when we gather together enough quanta to make a black hole, and no black hole actually forms in the true dual to the matrix model \cite{matrixmodel}. 

\b

To summarize, AdS/CFT gives us deep insights into the nature of gravitational solutions, but we cannot use the abstract idea of such a correspondence to argue away the information paradox. We may be able to use the duality map to learn something about gravitational solutions from their field theory duals, but in the end we must return to these gravity solutions to see how the Hawking argument is to be addressed.

 \subsection{Making black hole microstates}
 
 We now turn to the most interesting task: constructing actual black hole microstates.
 
 The  agreements of entropies (\ref{sixt}),(\ref{fift}) suggest that  bound states of branes should yield the states of a black hole. The estimate (\ref{thir}) of the size of these states suggests that these bound states may look nothing like the traditional black hole.  
 
\begin{figure}[htbt]
\hskip .5 in\includegraphics[scale=.45]{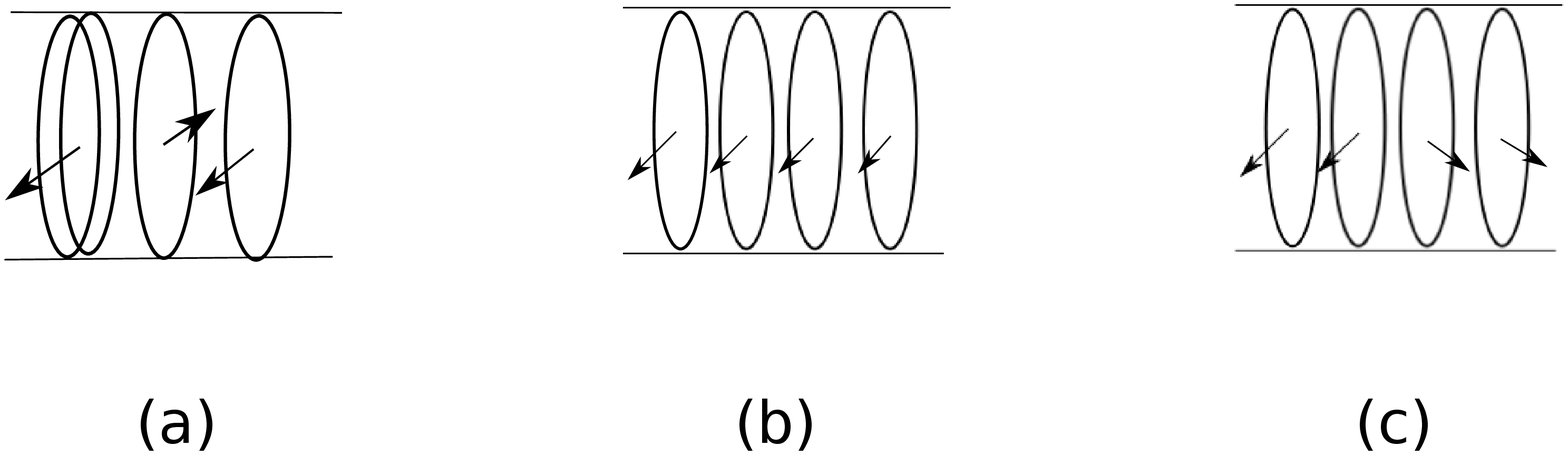}
%
%
\caption{ (a) The general state of the D1D5 extremal bound state is given by breaking the effective string into loops in all possible ways, with arbitrary directions for the fermion spins of the loops. (b) The simplest state of the D1D5 CFT has all loops of the effective string `singly wound', with all spins aligned. (b)  A state with all loops still `singly wound', but with spins such that the overall angular momentum vanishes. }
\label{fa2}       
\end{figure}

 It is good to start with a list of all the states that we wish to make. This list is best made by using the dual CFT. In the case of the D1D5 bound state the field theoretic description is given in terms of an `effective string' which has total winding number $n_1n_5$.  We can join this string into loops in different ways: for example we can have $n_1n_5$ separate loops each with winding number one, or a single loop with winding number $n_1n_5$, or any other combinations of loop winding numbers  which satisfy
 \be
 \sum_i N_i w_i =n_1n_5
 \ee
 where we have $N_i$ loops of winding number $w_i$ (see fig.\ref{fa2}(a)). Counting these possibilities yields the entropy $4\pi\sqrt{n_1n_5}$; here we have included an extra degeneracy that arises from 4 fermion zero modes on each loop.

 We can use this CFT description to  list all the the bound states  in order of `complexity'. The simplest state has all loops singly wound, and all fermion zero modes giving a spin in the same direction (fig.\ref{fa2}(b)). What could be the gravity solution corresponding to this microstate? 
 
 Since all the spins are aligned, the state has a large angular momentum $J$. It is natural to start by looking  for black hole solutions which have the mass and charge carried by $n_1$ D1 branes and $n_5$ D5 branes, and which have angular momentum $J$. It turns out that this value of $J$ makes the black hole `overrotate', so that the horizon area becomes {\it zero}; thus we get a regular geometry rather than a black hole \cite{bal,mm}. There is no horizon and no singularity. 
 
 Thus we have obtained a regular solution for this particular microstate.  This may not be considered very exciting, since it was well known that making a black hole `overrotate' would send its horizon area to zero, and  so it seems that we have just taken a very special state which does not correspond to a black hole. 
 So we now turn to the state in fig.\ref{fa2}(c), where all loops are still `singly wound' but the spins on half the loops cancel the spins on the other half. In this case $J=0$. We can look at  black hole solutions which again have the mass and charge of our D1, D5 branes, but now with $J=0$. In making these black hole solutions we must be careful  to  include the $\alpha'R^2$ term present in the string theory action in addition to the standard Einstein term $R$. Assuming a spherical symmetric ansatz and solving the supergravity equations we  find the usual black hole type solution with a horizon. So it seems we are back in the clutches of the Hawking paradox.  
 
 But it turns out that the true gravitational solution corresponding to this state with $J=0$ does {\it not} have spherical symmetry. In fact there is a completely regular solution, again with no horizon and no singularity, that describes the microstate in fig.\ref{fa2}(c). This breaking of symmetry is crucial to understand, so let us take a moment to outline its origin. 
 
 Recall that the D1D5 bound state can be mapped by dualities to a string carrying winding and momentum (eq.\ref{qwone})). This string is multiwound around a compact circle, but we can understand its vibration profile better by imagining it is opened up to a singly wound string on a circle with length $L_T$. Different vibration profiles of the string give all the states of the string, and by duality, all the states of the D1D5 system. 
 
 The maximally rotating D1D5 state of fig.\ref{fa2}(b) maps to a configuration where this string swings around making one turn of a uniform helix \cite{lm3}. This solution is axisymmetric, and so can be recovered by looking at the known class of black hole solutions and inserting the required quantum numbers. But the D1D5 state in fig.\ref{fa2}(c) maps to a string that swings clockwise on half its length and anticlockwise on the other half. Such a vibration has no spherical or axial symmetry, but it does map to a D1D5 solution with no horizon and no singularity, just as the maximally rotating helical string did \cite{lm4}. 
 
 To summarize, we have traditionally obtained the black hole geometry by starting with a spherically symmetric ansatz
 \be
 ds^2~=~-f(r) dt^2 + g(r) dr^2 +r^2 d\Omega^2 + dz_idz_i
 \label{qsone}
 \ee
 where we have assumed that the metric is independent of the angular variables and also independent any compact direction coordinates $z_i$. The actual black hole microstates in string theory are {\it not} spherically symmetric. Further,  the compact directions are involved in very nontrivial ways as we now explain.
 
 If the charges in our bound state are string winding and momentum, then  the geometry is generated  by the vibrating string source, and there is the expected `source singularity' at the location of the string. When we perform the dualities that map the charges to D1 and D5 branes, an interesting thing happens at the location of this source singularity. There still appears to be a singularity, but it is only a {\it coordinate} singularity \cite{lmm}. This coordinate singularity corresponds to a Kaluza-Klein monopole, which is generated by a nontrivial fibration of a compact circle over the noncompact directions. The net monopole charge of the entire solution vanishes, as it should because we have taken only D1 and D5 charges for our state. But the different allowed relative locations of the monopoles and antimonopoles gives a large phase space, which describes the entire set of D1D5 bound states.  In short, the compact directions twist in different ways at different values of the angular coordinates $\theta, \phi$, generating solutions that are regular everywhere but that have  no symmetries in general. We miss all such solutions if we start with the spherically symmetric ansatz (\ref{qsone}).

 Let us summarize the entire story that we learn from this setup of 2-charge extremal holes:

 \b

 (i) The number of extremal bound states of the D1D5 brane system can be counted by abstract topological methods, and give a microscopic entropy $S_{micro}=4\pi\sqrt{n_1n_5}$ \cite{sen}.
 
 (ii) If we assume a spherically symmetric ansatz and a trivial factorization of the compact directions, then the low energy supergravity action gives an extremal black hole with horizon, with a Bekenstein-Wald entropy $S_{bek}=4\pi\sqrt{n_1n_5}$ \cite{dabholkar}.
 
 (iii) The actual microstates of the D1D5 system can be constructed.  It is found that they are not spherically symmetric and the compact directions are locally nontrivially fibered, though the net monopole charge of these fibrations vanishes. The solutions have no horizon and no singularity \cite{lm4,skenderis}.
 
 (iv) The phase space of these horizonless gravitational solutions can be quantized, and yields the entropy $S=4\pi\sqrt{n_1n_5}$ \cite{rychkov}.
 
 (v) Though there is no horizon for any microstate, the  region where the typical microstates  exhibits their nontrivial structure has a boundary whose area $A$ satisfies $A/G\sim \sqrt{n_1n_5}\sim S_{bek}$ \cite{lm5}.
 
 \b

If we have such a picture for all holes, then the information problem is solved, since the traditional horizon has been completely corrupted, and  
Hawking's computation of entangled pairs becomes invalid.

We can now consider the more complicated hole made with 3 charges: D1, D5 and momentum, whose entropy is given by (\ref{fift}). This time it is not so easy to make all states in closed form, but large families of states have been constructed, largely due to a remarkable set of papers by Bena, Warner and their collaborators. In all cases the solutions have the same feature as seen for the 2-charge case: there is no horizon or singularity, spherical symmetry is broken, and the compact directions are fibered  nontrivially  over the non-compact directions \cite{fuzzball2,fuzzball3,fuzzball4}.  All the precise details of string theory go into the construction. For example, if we write the tension of the fundamental string  as $T=1/(2\pi\alpha')$, then string theory tells us that the Newton constant is related to this $\alpha'$ through $G_N=8\pi^6 g^2 \alpha'^4$
(here $g$ is the dimensionless string coupling). If we change a numerical factor in this relation, then the microstate solutions will either become singular, or acquire a horizon, or suffer closed timelike curves. But when all parameters have the correct values as given by string theory, the solutions have none of these pathological features.

  \subsection{Some common confusions about fuzzballs, and their clarification}\label{secq}

  We have arrived at a picture of black holes that is remarkably different from the traditional one. The estimate (\ref{thir}) suggested that bound states in string theory have a size that grows with the number of quanta in the state and also with the coupling $g$ of string theory. When the coupling vanishes, the size of the bound state may be string scale or  planck scale, but when we raise the coupling to the value where a black hole is expected then the bound state  wavefunctional is found to be nontrivial over a region of that is of horizon size. Our actual construction of microstates at large coupling indeed yielded solutions that were completely different from the traditional black hole geometry; in fact the microstates turned out to have no regular horizon. This is good in retrospect: if a microstate  had a horizon, then we would be led to attribute an entropy $S={A\over 4G}$ to this horizon. But  this would suggest that we are describing $Exp[S]$ microstates, which makes no sense if we are already talking about {\it one} microstate.
  
 In spite of the fact that this picture yields a good solution to the information problem, many people were unhappy with such a radical change to the traditional black hole. Their objections could be split into two rough categories:
 
 \b
 
 (a) The microstates that have been constructed correspond to very special cases. The 2-charge case is itself too simple a black hole, and the 3-charge states that have been constructed are too few to be representative of the general 3-charge microstate. The remaining states will be {\it traditional  black holes}. 
 
 (b) It is obvious that microstates should be fuzzballs. But this observation is trivial by itself. We have made fuzzballs starting with the simplest, but as we go towards generic microstates the gravitational solution will become very complicated  at short distances, with large quantum fluctuations. If we cannot accurately describe the fuzzballs in this generic case, perhaps we have learnt nothing at all. 
 
 \b
 
 Both of these criticisms have little merit, as they are based on an incorrect understanding of the paradox itself. But we learn something interesting by examining the errors involved, so we discuss them in more detail now. 
 
 \b

 (a') Suppose some states are fuzzballs but others are black holes. Let us make the definition of `black hole' very precise: In the traditional black hole we have a region around the horizon where the geometry is regular of the form (\ref{three}) in a `good slicing', and low energy physics in this region is just like  physics in the lab. But in this case the Hawking argument will lead to information loss or remnants, and neither seem acceptable in string theory. So what are the proponents of this view seeking?
 
 It turns out that this view arose with relativists, who were not using string theory as their theory of gravity, and who were comfortable with one of the models of remnants. The rationale for this view is clear. 
If we follow the collapse of a shell, it appears to leave behind it the traditional black hole geometry (\ref{three}). Why shouldn't we trust  this evolution, where every step seems to be in a valid semiclassical approximation?

With what we have learnt  about  fuzzballs, we can now postulate an answer. There is a very small but nonzero probability for the collapsing shell state to {\it tunnel} into one of the fuzzball states, since these fuzzball states have the same quantum numbers as the shell. We can estimate this tunneling amplitude as $Exp[-S_{cl}]$, where $S_{cl}$ is the Einstein action ${1\over 16\pi G}\int R \sqrt{-g}\, d^4 x$. To estimate  $S_{cl}$, we set all length scales to be $\sim GM$. Then $R\sim (GM)^{-2}$, and $\int \sqrt{-g} d^4 x\sim (GM)^4$, so we get
 \be
 S_{cl}\sim {1\over G } (GM)^{-2}(GM)^4\sim GM^2\sim \Big ({M\over m_{pl}}\Big )^2
 \label{qatwo}
 \ee
 Thus the tunneling amplitude is indeed very small, as expected for any tunneling process between two macroscopic configurations. But there are $Exp[S_{bek}]$ fuzzball states that we can tunnel to, so we must multiply the tunneling probability by the number of states ${\cal N}$, where
 \be
 {\cal N}=Exp[S_{bek}] = Exp[{A\over 4G}]=Exp[4\pi GM^2]\sim Exp[\Big ({M\over m_{pl}}\Big )^2]
 \label{qaone}
 \ee
 We see that the smallness of the tunneling amplitude can be compensated by the largeness of ${\cal N}$ \cite{tunnel}. The tunneling time was estimated in \cite{rate}, and was found to be much shorter than the Hawking evaporation time. Thus we find that the wavefunction of the collapsing shell will change to a linear combination of fuzzball wavefunctions, and the radiation process will then process from the surface of these fuzzballs just like radiation from any normal warm body.
    
 \b
 
 (b') The reason behind the second complaint was the following belief: ``The vacuum is a very complicated quantum mess at the planck scale. The generic fuzzball will be a complicated mess at the planck scale. Thus the generic fuzzball is practically indistinguishable from the vacuum for low energy physics; only planck scale processes would distinguish it from the vacuum. 
 Constructing special fuzzballs has been of no help; the information of the microstate is anyway expected to be distributed in planck scale details around the horizon, and we have obtained no usable description of this quantum mess for generic microstates''
 
  The logic behind this argument is flawed for two reasons. The first is just an erroneous understanding of the nature of states in quantum field theory. The vacuum indeed has violent fluctuations at the plank scale. But it is not correct to think that we can take a {\it different}  set of such fluctuations and obtain a state that is essentially similar to the vacuum at low energies. Assume for the moment  a Fock space description where all states are given  by operators $\hat a^\dagger_k$ acting on the vacuum $|0\rangle$. The only modifications we can make at the planck scale is by the action of $\hat a^\dagger_k$ with $k$ being planck scale. This adds an enormous energy density which would curve up the space into a planck sized ball. Thus it is not true that low energy physics can be left unaffected while we `hide information at the planck scale'.  
  
  The other error stemmed from the belief that one could leave the low energy dynamics unaltered at leading order and still get unitarity of black hole evaporation. This belief is simply false, as shown by the inequality (\ref{seven}).

  \b
  
  As we will see next, the fuzzball construction has not only allowed us to understand how we can have nontrivial structure at the horizon; we can also explicitly see radiation emerging from simple fuzzballs by a unitary radiation process.

\subsection{Hawking radiation from fuzzballs} \label{secrad}

As with most efforts in string theory, the first attempts at making microstates  were focused on extremal states -- ones with mass equal to charge. Such states have no extra energy to radiate, and the temperature of their corresponding black holes is zero. But soon nonextremal states were constructed, and these do radiate. How does this radiation process differ from the pair creation process found by Hawking, which leads to the information paradox?

Let us first describe how we compute emission from a CFT state.
The 3 charge non-extremal D1-D5-momentum states were described in fig.\ref{fa5}(b): we have an effective string with total winding number $n_1n_5$ carrying both left and right moving vibrations. The set of allowed states of the system are given by letting the effective string break up  into loops of arbitrary length, and letting the vibrations distribute themselves in arbitrary ways on these loops. The emission rate is found by computing the interaction vertex $V$ that couples a left moving vibration, a right moving vibration, and an emitted graviton. We then multiply this vertex $V$ with the occupation numbers $\rho_L, \rho_R$ of the left and right moving vibrations, to get the radiation rate
from this microscopic computation
\be
\Gamma_{micro}=\rho_L\rho_R V
\label{qwten}
\ee
If we take generic distributions for the vibrations given by Bose and Fermi distributions $\rho_B, \rho_F$ for bosonic and fermionic vibrations, then we reproduce the Hawing radiation {\it rate} computed from the traditional black hole geometry (eq.(\ref{tw})). 

As we have noted before, the radiation from the effective string is a normal unitary quantum process, while that from the hole leads to the information paradox. But we have now learnt that the black hole geometry is incorrect; we should take a fuzzball solution instead and then perform the gravity computation. 

\begin{figure}[htbt]
\hskip .5 in\includegraphics[scale=.45]{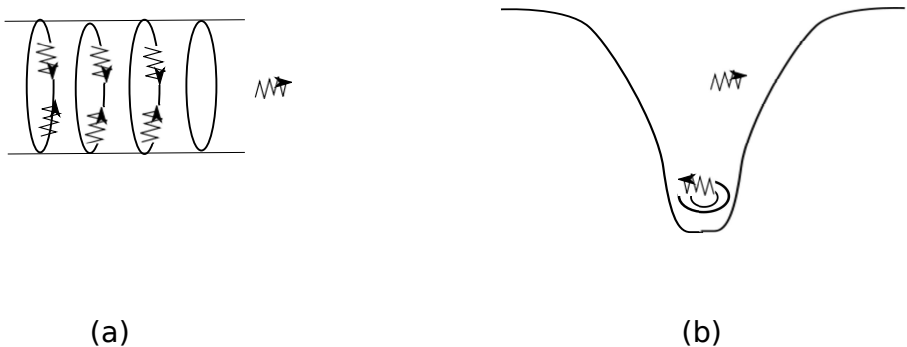}
%
%
\caption{ (a) The simplest nonextremal CFT state has all loops singly wound with the same excitations on each loop; the collision of oppositely moving vibrations gives an emission rate $\Gamma_{micro}$. (b) The corresponding gravitation solution is a `fuzzball' with no horizon. There is an ergoregion however (the doughnut shaped region in the interior of the geometry), and the rate of ergoregion emission $\Gamma_{gravity}$ agrees with $\Gamma_{micro}$. }
\label{fa6}       
\end{figure}

To do such a computation, we start with the simplest nonextremal CFT state, depicted in fig.\ref{fa6}(a). All loops of the effective string are `singly wound', and all loops carry the same set of vibrations. Let these vibrations be carried by fermionic modes which have their spins aligned and which are  packed in to fill their fermi level. 
 Let this distribution of excitations be denoted by $\bar \rho_L, \bar \rho_R$.  Using (\ref{qwten}),  the microscopic computation of emission from this microstate is given by
 \be
 \bar\Gamma_{micro}=V\bar\rho_L\bar \rho_R
 \ee
 The gravitational solution describing the microstate of fig.\ref{fa6} has been constructed \cite{ross}. It again has no horizon or singularity. Since there is no horizon, we cannot get emission from Hawking's process.  What we find instead is that the geometry has an {\it ergoregion}: a region where the Killing vector that is timelike at infinity becomes {\it spacelike}. In such a situation we cannot make the usual foliation of spacetime by spacelike slices and use the Killing vector for defining `time evolution'. Any time evolution must be defined using a foliation that `changes from slice to slice', and as a consequence we again get particle production. Such particle production from ergoregions has of course been known for a long time, and in fact was a precursor to Hawking's computation of emission from black hole horizons. What is important here is that  ergoregion emission  does {\it not} lead to an information problem: one member of the created pair floats to infinity as radiation, while the other settles down in the ergoregion, where it can influence the production of later pairs and thus transfer its `information' to them. Such an information transfer happens of course in radiation from all normal bodies, and we see now that the fuzzball solution for our microstate behaves just like any normal body, rather than like a black hole. The ergoregion emission rate was computed in \cite{myers}. We find that this rate $\Gamma_{gravity}$  agrees exactly with the expected rate of radiation from the microstate \cite{cm1}
\be
\bar\Gamma_{micro}~=~\Gamma_{gravity}
\label{qwel}
\ee

What we have seen here is very interesting.  We have looked at all the states of the hole as described by the CFT, and picked the simplest one to start with. The gravitational solution for this state has been explicitly constructed, and is found to have no horizon or singularity. The solution does have an ergoregion, which gives emission, but by a process that does not lead to information loss. The {\it rate} of this emission is however exactly the Hawking emission rate that we would have expected for this particular microstate, in the following sense.  The relation  (\ref{qwel}) shows that the ergoregion emission agrees with the emission computed in the CFT. Such CFT emission in turn can be computed for all CFT states, and when applied to generic states reproduces the emission rate found by Hawking (eq.(\ref{qwten})). Thus we can imagine that when we move to more generic states (from the special state in fig.\ref{fa6}), the fuzzball solution will become more complicated, and its emission will tend to the Hawking emission spectrum.  But now the emission will be information preserving just like radiation from any normal warm body.

\section{Conjectures}\label{sfour}

In the above section we have found that our picture of the black hole needs to be modified in a very significant way: instead of a vacuum at the horizon we find `fuzzball structure'. What we will now see is that this modification suggests equally significant changes in our picture of spacetime itself. 

We have not constructed all fuzzball states for all black holes; in fact very few nonextremal states have been constructed. What we will do is extrapolate our understanding of the known fuzzball states to obtain a qualitative picture that might work for {\it all} black hole microstates. We will then use this picture to conjecture solutions to many long standing questions about the nature of spacetime and entropy.

\subsection{A qualitative picture of general fuzzballs}

Consider the traditional Schwarzschild hole, with metric (\ref{two}). This geometry can be continued smoothly inside the horizon $r=2M$. By contrast,  fuzzball solutions `end' before a horizon is reached. This `end' does not signify that we have a sharp edge to our geometry.  Rather, a compact direction pinches off to zero size, creating a smooth `cap'.  Thus there is no sense in which we can continue `further inside'. This structure is reminiscent of the `bubble of nothing' \cite{witten} where a compact directions shrinks and `closes off space'. But in our case the compact circle is not orthogonal to the noncompact directions, and the structure we get instead is that of  Kaluza-Klein monopoles and antimonopoles.  Note that  while the `bubble of nothing' has a singular boundary condition for fermions at the place where the circle pinches, the KK monopole has the fibration structure required to give a regular boundary condition.  The KK monopole is one of the basic charges of string theory, and can be mapped by dualities to any of the other D-brane charges.  
The general microstate has not only monopoles but other string charges as well, so it may be best to just say that the gravitational state `ends' in a structure which contains the  allowed sources of  string theory.
The number of such states is expected to reproduce the Bekenstein entropy
\be
S~=~\ln {\cal N}~=~{A\over 4G}
\ee
A final point about the size of the typical fuzzball solution. Making a  KK monopole costs energy, and we should note that our total energy budget for the state is $M$, the mass of the hole we seek to describe. The simplest fuzzball states have only a few monopoles and antimonopoles, and their size is in general significantly bigger than the Schwarzschild radius $2M$. Since the number of monopole pairs is small, there is a limited amount of entropy in such solutions. To be able to make more monopole pairs, we can let the monopoles cluster closer to the radius $2M$; now we have a gravitational redshift that  allows more monopoles to be created for the same total mass $M$. Moving closer and closer to $r=2M$ we can get more and more effective phase space, until quantum fluctuations cut off this phase space.  Thus the generic fuzzball is expected to have  its structure at $r=2M+\epsilon$, where $\epsilon$ is small, perhaps planck scale. 

This is a very rough picture of fuzzball microstates, but we will see that all we need is the fact that such a picture {\it exists}; we do not need the detailed structure of the fuzzballs. We will put this picture of black hole microstates in the context of ideas suggested by Israel \cite{israel}, Maldacena \cite{eternal} and Van Raamsdonk \cite{raamsdonk} to arrive at our  picture of spacetime structure.

\subsection{The entropy of Rindler space}

\begin{figure}[htbp]
\begin{center}
\includegraphics[scale=.65]{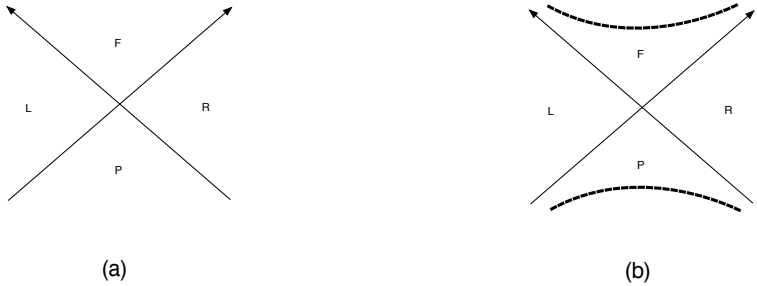}
\caption{{(a) Minkowski space and its Rindler quadrants (Right, Left, Forward and Past). (b) The Penrose diagram of the extended Schwarzschild hole. The region near the intersection of horizons is similar in the two cases.}}
\label{fn5}
\end{center}
\end{figure}

One question that has always puzzled relativists is the following. If black holes have an entropy given by their horizon area, should we associate an entropy with {\it all} horizons? In particular we can take empty Minowski spacetime, and choose `Rindler coordinates' that cover only the `right' quadrant. In these coordinates we see horizons at the boundary of this quadrant (fig.\ref{fn5}(a)). This region of Minkowski space looks very similar to the central region of the full black hole Penrose diagram, fig.\ref{fn5}(b). Should we associate an entropy 
\be
S_{rindler}={A\over 4G}
\label{qone}
\ee
to any area $A$ of  the Rindler horizon? 

For a black hole we could think of the entropy as arising from the number of ways we could make the hole. But in the Rindler case, what would such an entropy be counting? 
Even though the answer to this was not clear, the expression (\ref{qone}) was generally accepted as holding for Rindler horizons, and in particular was used to `derive' Einstein's equation from thermodynamics \cite{jacobson}. What we will now see is that in terms of fuzzballs, there is a logical explanation of (\ref{qone}) as a count of states, even though we are describing the Rindler quadrants of {\it empty} Minkowski space \cite{plumberg}.

Consider the spacelike slice  $z=0$ in the Minkowski spacetime of fig.\ref{fn5}(a). Consider a free scalar field $\phi$, and let $|0\rangle_M$ be the vacuum state of this scalar field on Minkowski space. 
 Half of our spacelike slice $z=0$ lies in the left Rindler wedge and half in the right. We can write the complete state $|0\rangle_M$ in terms of states in the left and right wedges
  \be
|0\rangle_M=C\sum_i e^{-{E_i\over 2}}|E_i\rangle_L|E_i\rangle_R, ~~~~~~~C=\Big (\sum_i e^{-E_i}\Big )^{-\h}
\label{split}
\ee
Not surprisingly, the state $|0\rangle_M$ is  entangled between the left and right Rindler wedges. The states $|E_i\rangle_L, |E_i\rangle_R$ are energy eigenstates under the `Rindler time' coordinate in each wedge. For the free scalar field, these eigenstates are explicitly known, though the sum in (\ref{split}) needs regularization at $E_i\r\infty$.

We now make a few observations. First, if we had an interacting scalar field, the states $|E_i\rangle_L, |E_i\rangle_R$ would be eigenstates of the {\it interacting} Hamiltonian. Second, an expansion like (\ref{split}) is expected for {\it any} field, and one field that is always present is nature is the graviton field $h_{ij}$. Let us therefore ask the question that will be central for us: what is the analogue of (\ref{split}) for gravitational fluctuations $h_{ij}$? 

Note that the strength of gravitational interactions increases with energy.  The states $E_i$ have large local energy density  in the region close to the Rindler horizons, due to the large redshift near the horizons in the Rindler metric. Thus the states $|E_i\rangle_R$ for the gravitational field are expected  to be states with the following characteristics: (i) they should `live' in only the right Rindler quadrant (ii) they will have high local energy density near the Rindler horizons (iii) they will involve very nonlinear gravitational interactions near the horizons.

\begin{figure}[htbp]
\begin{center}
\hskip -1 in \includegraphics[scale=.60]{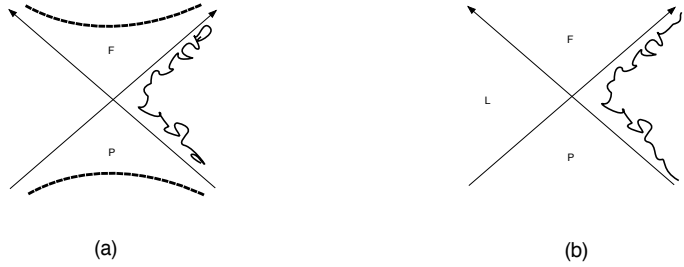}
 \caption{{(a) The eternal black hole spacetime. The geometry of a fuzzball microstate is only the region to the right of the jagged line, and so it lies only in the right quadrant. (b) Taking the limit $M\r \infty$ we obtain fuzzball states lying in the right quadrant of Minlowski spacetime.}}
\label{fn11p}
\end{center}
\end{figure}

\begin{figure}[htbp]
\begin{center}
 \includegraphics[scale=.65]{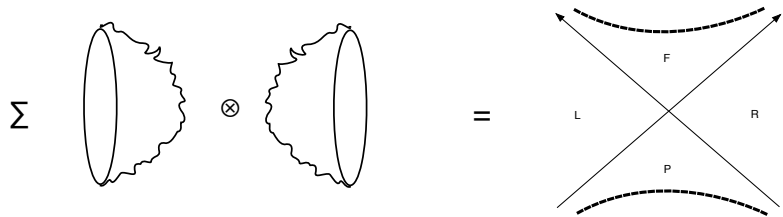}
\caption{{Black hole microstates are fuzzballs that `end' without forming a horizon. Summing over pairs of microstates (with appropriate weights) should give the geometry of the extended Schwarzschild hole.}}
\label{fn7}
\end{center}
\end{figure}

But these are just the characteristics of the fuzzball solutions that have been found! The fuzzballs end just outside the the place where the horizon would have occurred. We depict this in fig.\ref{fn11p}(a), where we draw the eternal black hole diagram, and then indicate the boundary of the fuzzball as a jagged line in the right quadrant; thus the fuzzball geometry is only the region to the right of this jagged line. The fuzzball structure in the vicinity of this jagged line  is over very short length scales and very nonperturbative: we have a complicated distribution of  monoples and antimonopoles.  Taking the limit $M\r\infty$ of the black hole mass brings us to Minkowski space, (fig.\ref{fn5}). Thus it is natural to conjecture that the fuzzballs obtained in this limit  are just the solutions $|E_i\rangle_R$ appearing in the decomposition of the Minkowski vacuum $|0\rangle_M$ for the gravitational field. Following general ideas of Van Raamsdonk \cite{raamsdonk}, we depict the relation (\ref{split}) for the graviton field pictorially in fig.\ref{fn7}. 

We now have a natural conjecture for the meaning of (\ref{qone}). Consider 3+1 Minkowski spacetime, but as a part of full string theory, so that we have 6 compact directions, and the branes etc.\ that are the allowed sources in string theory. Further, consider this Minkowski spacetime as the limit of the central part of an eternal  black hole diagram in the limit $M\r\infty$. 
There will exist gravitational solutions that approach flat space near infinity, but which end in a monopole-antimonpole structure near the Rindler horizons, such that spacetime `ends' before reaching these horizons.  (We will of course also have  the other branes/fluxes around these monopoles needed to obtain the details of the full fuzzball structure). In the 2-charge extremal case the space of such solutions was quantized and found to yield the correct entropy. Though we cannot yet construct the uncharged solutions needed for the Rindler case, it is natural to conjecture that quantizing the space of solutions will yield (\ref{qone}). 

Thus, roughly speaking,  Rindler entropy counts the number of  manifolds without boundary that fill the right Rindler wedge.

\subsection{Relation to earlier ideas}

Let us now connect our above discussion to ideas that have been presented elsewhere. The Penrose diagram of the `eternal hole' of fig.\ref{fn5}(b) has two asymptotic infinities, rather than the single infinity that 
we normally have in our world. Israel \cite{israel} postulated that we should think of the two sides of this Penrose digram as describing two copies of our gravitational physics, in the same way that we take two copies of a field theory when using the `real time formalism' to study finite temperature  dynamics. Maldacena \cite{eternal} studied the dual CFT description of the eternal hole in AdS space.  This time there are two asymptotically AdS boundaries, and we should associate a CFT with each boundary. Using Israel's connection to the real time formalism, Maldacena arrived at the conclusion that the CFT state describing the eternal hole  is an {\it entangled} state  of the form
\be
|\psi\rangle=\sum_k e^{-{E_k\over 2T}}|E_k\rangle_L\otimes |E_k\rangle_R
\ee
where the two copies of the CFT from the two sides of the hole are seen to be entangled.

Van Raamsdonk has recently taken this notion of entanglement further, to the entanglement of {\it gravitational} solutions. Consider one of the two copies of the CFT in Maldacena's description. Each state $|E_k\rangle_L$ should be dual to some gravitational solution $|g_k\rangle_L$, and similarly $|E_k\rangle_R$ should be dual to a gravitational solution $|g_k\rangle_R$. Thus we should be able to write the eternal black hole geometry as an entangled sum of {\it gravitational} solutions 
\be
|g\rangle_{eternal}=\sum_k e^{-{E_k\over 2T}}|g_k\rangle_L\otimes |g_k\rangle_R
\label{qwfourt}
\ee
This is interesting, since the spacetime on the  LHS is a  geometry that is {\it connected}  between its left and right sides, while the gravitational solutions appearing on the RHS have {\it no} connection between the L and R sets. Thus we conclude that if we take disconnected gravitational manifolds, but entangle their states, then we generate a connection between the manifolds. 

\begin{figure}[htbt]
\hskip .5 in\includegraphics[scale=.25]{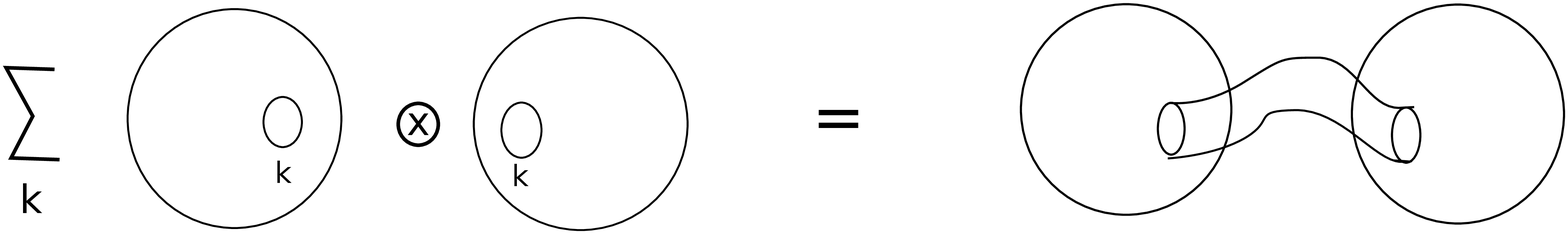}
%
%
\caption{The `sewing prescription in 2-d CFTs.  Summing over states $\psi_k$ in two disconnected spheres gives the CFT on  a connected manifold.}
\label{fa11}       
\end{figure}

 While this notion may appear strange at first, we note that  a simple version of this effect can be seen in a more familiar context: the idea of `sewing' in 2-d CFTs. Consider a 2-d CFT, say the Ising model, on a sphere. Cut a hole on this sphere, and on the boundary of this hole put a state $|\psi_k\rangle$. Take a second sphere, cut a similar hole, and put the same state  $|\psi_k\rangle$ on this hole. The two spheres are not connected, but supposing we perform the sum
 \be
\sum_k  e^{-\tau E_k} |\psi_k\rangle_1\otimes |\psi_k\rangle_2
\label{qwfift}
\ee
then we `sew' the two spheres together, generating a smooth handle joining the hole on the first sphere to the hole on the second sphere. Thus `entangling' the two spheres creates a space that {\it connects} the spheres (fig.\ref{fa11}). 

Returning to the postulate (\ref{qwfourt}), we do notice a potential difficulty. In a theory of gravity, most of the states at an energy $E$ are expected to be {\it black holes}. What geometry $g$ should we take for such states? If we take the traditional metric with horizon, then this metric can be continued past the horizon, and into another asymptotically AdS region; thus our state $|g_k\rangle_L$ which was supposed to describe a metric with {\it one boundary} now seems to  describe a metric with {\it two} asymptotic  boundaries. But this is where our understanding of black hole microstates helps us.  These microstates are fuzzballs, with no horizon, and there is no distinction in principle between black hole states and `normal' states.  Thus the sum (\ref{qwfourt}) does make sense. Extending this notion to asymptotically flat space solutions (instead of asymptotically AdS) gives us the postulate depicted in fig.\ref{fn7}. Taking the limit $M\r \infty$ for the black hole brings us to Rindler space, and the corresponding interpretation of Rindler entropy as arising from fuzzball solutions entangled between the left and right wedges. 

 \begin{figure}[htbt]
\hskip .5 in\includegraphics[scale=.45]{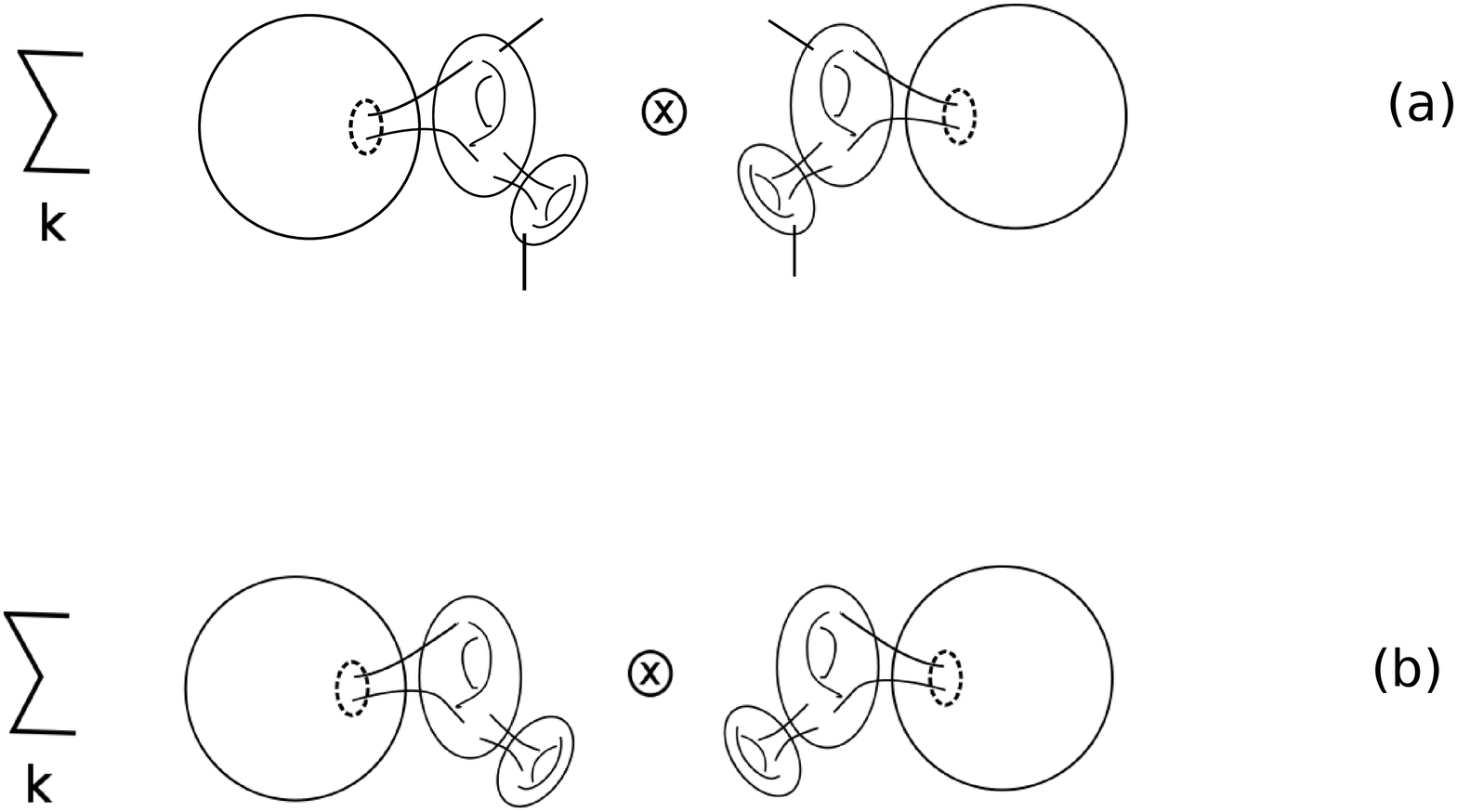}
%
%
\caption{(a) Replacing the CFT states in fig.\ref{fa11} by path integrals over manifolds with operator insertions. (b) In the gravity case we have a similar picture, but there are no insertions.}
\label{fa12}       
\end{figure}

A final comment on the relation between (\ref{qwfourt}) and (\ref{qwfift}). Consider one of the states $|\psi_k\rangle_1$ in the CFT case (\ref{qwfift}). This state is defined at the boundary of a circle, but we could try to continue the state inside the circle. Some states $|\psi_k\rangle_1$ can be obtained by a CFT path integral over a higher genus manifold; in this case we can `fill in'  the circle by this compact manifold (fig.\ref{fa12}(a)). But in general we will also need to add insertions of operators. For example if the CFT was the Ising model, we would need insertions of the spin operator $\sigma$ to get the most general state $|\psi_k\rangle_1$, since the path integral on a compact manifold cannot generate states with nonzero expectation value of the spin. We depict such insertions by external legs in fig.\ref{fa12}(a). 

Now consider the relation (\ref{qwfourt}) for the case of gravity. Fuzzball solutions are found to `end smoothly' without horizon.  This situation is analogous to that for the CFT states which could be obtained by a path integral over a compact manifold {\it without} extra insertions. The difference is that in the gravitational case  we expect {\it all} states to have this form. Of course the fuzzball solutions contain the strings, branes etc.\ that are valid sources in the theory, but we regard these as part of the `capped off' gravitational solution. In short, string theory is a `complete' theory, and this seems to be reflected in the fact that all its states  `cap off' by themselves without the need for extra insertions (fig.\ref{fa12}(b)).

\subsection{Entropy of de-Sitter space}

\begin{figure}[htbt]
\hskip .5 in\includegraphics[scale=.45]{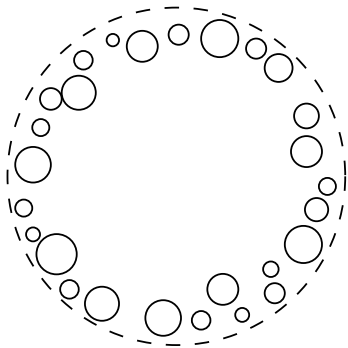}
%
%
\caption{ The horizon of a static patch of de Sitter is indicated by the dashed line. The collection of bubbles indicates a `fuzzball solution' which is de Sitter in its interior but which `ends' in a set of monopole-antimonopole solutions near the horizon. The count of such solutions is conjectured to give the entropy of de Sitter space.  }
\label{fn8}       
\end{figure}

Let us now turn to another question that has been of interest: how should we understand the entropy of de Sitter space? This spacetime is expanding due to the presence of a positive cosmological constant.  We can choose coordinates in which the metric appears static, and then we get a horizon with an area $A$ at the boundary of this static patch. But this boundary can be moved around depending on our choice of static coordinates, so it was  unclear if we should attach any  meaning  to the entropy
\be
S_{de-Sitter}={A\over 4G}
\label{qtwo}
\ee
But we can now understand this entropy in just the way we understood  Rindler entropy. The complete state straddling both sides of the horizon can be written as an entangled  sum of states on each side. The entropy of de Sitter gives the count of all gravitational solutions that approach regular de Sitter at the center of the static patch, but that end without boundary in a structure of monopoles and antimonopoles before reaching the boundary of the static patch. Thus the entropy (\ref{qtwo})  counts compact manifolds without boundary for the situation where we have a positive cosmological constant (fig.\ref{fn8}).

\subsection{Complementarity}

One of the most polarizing ideas concerning the information paradox has been the notion of {\it complementarity}, first put forth by 't Hooft and then developed in detail by Susskind and others \cite{complementarity}. This idea seems to bypass Hawking's paradox, but a closer look reveals that in ordinary gravity theories the idea is {\it wrong}. What we will see now is that once the black hole microstates are known to be fuzzballs, there is indeed a version of complementarity that we can extract.

Roughly speaking, the idea of complementarity is the following.  The trouble with the black hole was that information fell in through the horizon and disappeared, instead of showing up in the Hawking radiation. We can avoid this problem if we assume that the region outside the horizon $r>2M$ is somehow complete by itself, so nothing need ever be thought of as falling in through the horizon.  We imagine a membrane placed at a `stretched horizon' at $r=2M+l_p$ which absorbs infalling matter and radiates it back, unitarily, as Hawking radiation. 

The motivation for this picture came from the fact that the Schwarzschild coordinates (\ref{two}) fail at the horizon. There are two consequences of this failure:

 (i) The Schwarzschild time coordinate $t$ tracks infalling quanta only down to the horizon but not past it; thus these quanta appear to slow down and accumulate just outside $r=2M$
 
  (ii) Due to the large redshift near $r=2M$, the quantum fluctuations of the geometry are large on $t=constant$ slices.  
  
  These two facts suggested the possibility that strong quantum gravity effects modify physics in a way that make the region $r>2M$ a complete domain for physics, and since nothing would then fall past the horizon, there would be no information problem to worry about.

There is of course an immediate objection to this line of reasoning. The Schwarzschild coordinates fail at the horizon, so instead of drawing conclusions from these coordinates we should look for {\it good} coordinates like (\ref{three}) that take us smoothly across the horizon.  And in these good coordinates we see no violent quantum gravitational fluctuations that would cause physics to end at the horizon, or any evidence that matter stays confined to  $r>2M$. And if matter can fall into the region $r<2M$ in good coordinates, then don't we need some role for this region too?

 `Complementarity' gets around this problem by conjecturing that the region $r<2M$ gives a {\it dual} description of  physics occurring at $r>2M$. That is, the interior of the black hole gives another copy of  the physics that we have in the exterior of the hole. We do not get a contradiction from such a duplication of information because anything in the region $r<2M$ lives only a short time before getting destroyed at the singularity; thus there is not enough time to do a detailed comparison of the inside and outside copies.

While this picture of `complementarity' appears to bypass the information problem, it is unclear how we can reconcile it with  the usual idea of Hamiltonian evolution on Cauchy slices. Suppose we take the good coordinates (\ref{three}) where a complete Cauchy surface has parts both inside and outside the horizon. Is complementarity asking us to keep only the Hilbert space outside $r=2M$? If so, this would be strange, since there seems to be no special role to the point $r=2M$ on a `good slice' through the horizon. If we follow Hamiltonian evolution to a later Cauchy slice, we see Hawking's pair creation and the consequent entanglement between the inside and the outside. How can we bypass the information problem that follows from this entanglement?

The fuzzball picture of black hole microstates completely changes our perspective on what the black hole looks like. With this picture of microstates,  we will see that there {\it is} a version of complementarity that we can extract:

\b

(a) The first part of the idea of complementarity  is in fact immediate. We have found that black hole microstates do not have horizons. Collapse of a shell gives a pure state that, after the tunneling process discussed in section \ref{secq}, becomes a complicated gravitational solution ending in monopoles-antimonopoles near $r=2M$. 
Thus instead of an imaginary membrane at a `stretched horizon' \cite{membranebook}
we have a real `membrane' that can absorb and re-radiate energy \cite{membrane}. Note that it is not the badness of coordinates that gives us the barrier near $r=2M$; the spacetime structure really `ends' there when the compact directions pinch off.  We can see this structure and the resulting  absorption and radiation of quanta explicitly for the simple microstates discussed in section \ref{secrad}, and we can imagine that a similar picture holds for more general microstates as well. In short, regular  spacetime does end near the location where the horizon would have been in the traditional geometry.

\b

(b) The above observation  gets us only half of the complementarity picture; we still have to ask if there is any sense in which the interior of the horizon can be recovered in a `dual description'.  The horizon structure of the eternal hole is similar to the structure of Rindler horizons in Minkowski space (fig.\ref{fn5}), so let us map the black hole problem to the corresponding problem in Rindler space. Observers outside the black hole horizon are described by operators $O_R$ which, in the Minkowski problem, will be located in the right Rindler wedge. A hole made by collapse is in a definite pure state, which we can take to be the analogue of one of the Rindler states  $|E_k\rangle$. Thus measurements outside the hole correspond to Rindler correlators  ${}_R\langle E_k|\hat O_R|E_k\rangle_R$. There is no evidence of horizon-like behavior so far, since a Rindler state ends in a complicated mess before reaching the location of the horizon.

 The {\it full} Minkowski spacetime does have the Rindler horizons. Suppose we compute the correlator of the same operator $\hat O_R$  as above but in the full Minkowski vacuum; this would be the analogue of computing the correlator in the eternal black hole spacetime with horizons. Noting the decomposition (\ref{split}), we find
\bea
{}_M\langle 0|\hat O_R|0\rangle_M&=&C^2\sum_{i,j}e^{-{E_i\over 2}}e^{-{E_j\over 2}}{}_L\langle E_i|E_j\rangle_L {}_R\langle E_i|\hat O_R|E_j\rangle_R\nn
&=&C^2\sum_i e^{-E_i}{}_R\langle E_i|\hat O_R|E_i\rangle_R
\label{qwe1}
\eea
Thus the expectation value in the Minkowski vacuum is given by  a thermal average over the Rindler states. By contrast, our physical situation needed us to compute the correlator in {\it one} Rindler state: ${}_R\langle E_k|\hat O_R|E_k\rangle_R$.  

  Now we come to a crucial point, which is a basic fact of statistical mechanics:  for a {\it generic} state $|E_k\rangle_R$ and {\it appropriate} operators $\hat O_R$ we should be able to replace expectation values in the state $|E_k\rangle_R$ by an ensemble average
\be
{}_R\langle E_k|\hat O_R|E_k\rangle_R\approx {1\over \sum_l e^{-E_l}}\sum_i e^{-E_i}{}_R\langle E_i|\hat O_R|E_i\rangle_R={}_M\langle 0|\hat O_R|0\rangle_M
\label{qwe2}
\ee
For example if we stick a thermometer in a beaker of water, then the rise of mercury can be computed using the actual state $|\psi_k\rangle$ of the water, {\it or} by using the ensemble average over such states; the result is expected to be the same to leading order. Here the state of water is assumed to be a {\it generic} state, and the operator measuring temperature is of the  `appropriate' type mentioned above. If on the other hand we had tried to measure the Brownian motion of one water molecule in the beaker, the result would depend very sensitively on which $|\psi_k\rangle$ we took, and we would {\it not} have the approximate equality (\ref{qwe2}).

In short, even though a single fuzzball has no region which is `inside the horizon', we  still find that the expectation value of `appropriate operators'  can be computed to good accuracy by using the {\it traditional} eternal black hole geometry which extends smoothly past a regular horizon. Thus we have obtained an analogue of complementarity,  with two small caveats: (i) it is not a general relation for all operators, but for operators that have an appropriate thermodynamic behavior (ii) it is at its heart an {\it approximate} relation, though the approximation (\ref{qwe2}) would be excellent for large black holes.

\b

In what follows we will discuss the nature of the `appropriate' operators in more detail.

\subsection{The infall problem}

A central question with black holes has always been: what happens when something falls through the horizon? 

The traditional geometry (\ref{two}) gives a simple solution to the Einstein equations, and appears robust due to the `no hair' theorems.  This geometry led to a very strong prejudice in the field that  {\it nothing} happens at the horizon; we just find a region of vacuum spacetime. But the information paradox tells us that at least for the low energy modes $(E\sim kT$) involved in Hawking radiation, we should find corrections of order {\it unity}. Thus if we are to preserve the traditional geometry (\ref{two}) in any sense, all we can hope for is that it gives an effective behavior for {\it high} energy modes $E\gg kT$.

We have seen that individual black hole microstates have nontrivial structure at the horizon which allows them to emit low energy Hawking radiation by an information conserving process. It is tempting to think that the `fuzzball' is a very fuzzy low density object, so that a high energy quantum with $E\gg kT$  moves through it much as it would move through the vacuum. In that case we could say that the traditional black hole geometry is still the metric seen by infalling objects, since such objects are blueshifted to high energy by the time they reach the horizon.\footnote{Some authors have suggested that the traditional black hole geometry can be obtained by averaging over microstates in suitable ways \cite{bala}.}

While it is possible that such will turn out to be the case when we know more about fuzzballs, there is a completely different interpretation of infall dynamics that is suggested by the behavior of the fuzzballs that {\it have} been constructed. It does not appear that a high energy quantum can move in a straight line through the monopole-antimonople structure
of the fuzzball. Instead, it appears that the quantum will impact the surface of the fuzzball, and  transfer its energy to the fuzzball. What happens then?

Before conjecturing an answer, let us note that the question we are now probing is, in a sense, the {\it opposite} of the information problem. For the information paradox, we needed to find a way in which different microstates would behave {\it differently}, so that the information of the microstate could be encoded in outgoing radiation. For the `infall problem', we wish to ask if generic microstates can somehow behave the {\it same}, when hit by high energy quanta.

In this context it is natural to conjecture that the high energy ($E\gg kT$) quanta falling onto the fuzzball excite {\it collective} modes of the fuzzball.   The spectrum of such collective modes is expected to be independent of the precise choice of microstate, assuming that we choose a generic microstate. 

So far the situation is not different from that of a drop of water: hitting the drop with an energetic particle causes collective oscillations that are insensitive to the exact atomic arrangement in the drop. But now we will add a further step to our conjecture. Let the Hilbert space ${\cal H}$ of quanta outside the fuzzball be given by states $|\psi_k\rangle$, with energies $E_k$. Let the state $|\psi_k\rangle$, upon impacting the fuzzball, excite a collective mode $|\t\psi_k\rangle$, which has energy $\t E_k \approx E_k$. Thus the Hilbert space of infalling quanta maps to a virtually isomorphic Hilbert space $\t {\cal H}$ of collective excitations of the fuzzball. The question we ask is: in this case, can we `know' that quantum has impacted the fuzzball?

A little thought tells us that in this situation we {\it cannot} tell that the quantum has been absorbed by the fuzzball. Quantum mechanics keeps track of the relative motion of Hilbert space vectors, but if we map a Hilbert space into a different one while preserving all norms and inner products, then the dynamics has remained unaltered. Thus we seem to have kept our cake and eaten it too: the quantum did not penetrate into the fuzzball, but it left a virtually faithful copy of itself on the collective modes of the fuzzball. This is very reminiscent of the membrane that one 
might have imagined being placed at the `stretched horizon', but instead of being a virtual construct, the membrane is now real, being made up of the degrees of freedom of the fuzzball microstate. 

What we must do now is offer some justification for the above picture. For this we recall the relation (\ref{qwe2}). The LHS gives the expectation value of an operator in a Rindler state, which corresponds to a fuzzball state of the black hole. The RHS corresponds to an expectation value in the Minkowski vacuum, which corresponds to the eternal black hole. The question is: for which operators $\hat O_R$ is the approximation (\ref{qwe2}) good? We postulate that {\it these operators are the ones that describe the physics of infalling quanta with $E\gg kT$}.  

With this postulate, we have what we want. Let the operator $\hat O_R$ correspond to a measurement appropriate to an infalling object ($E\gg kT$). The object hits the fuzzball and excites collective oscillations, getting an expectation value given by the LHS of (\ref{qwe2}). But this is equal, to a good approximation, to the expectation value of the RHS of (\ref{qwe2}). This latter expectation value {\it corresponds to the same measurement made in the eternal black hole spacetime which has no structure at the horizon, just smooth spacetime}. Note that in obtaining this equivalence we have made use of the  the identity pictured in fig.\ref{fn7}. 

Finally, a small comment on the nature of $E\gg kT$ operators. The energy levels of the fuzzball are closely spaced,  since the entropy of the hole is so large. When a quantum with energy $E$ comes near the surface of the fuzzball with mass $M$, the tunneling process discussed in section \ref{secq} causes the wavefunction to spread over all fuzzball states with total  energy $M+E$. Using the relation $TdS=dE$, we see that the change in entropy is  $dS=dE/T\gg 1$ for quanta with energy $E\gg kT$. Since entropy is the log of the volume of phase space, we see that the phase space that is  accessible with energy $M+E$  is much larger than  the original phase space at energy $M$. The absorption process is thus like a `Fermi golden rule' absorption, where the incoming quantum transitions into a closely spaced band of levels in the fuzzball. In this situation the precise initial state of the fuzzball is not relevant; the dynamics is governed by the level density and the average coupling to the fuzzball states. This offers a rationale for why it is the $E\gg kT$ operators that satisfy the approximation (\ref{qwe2}) in which the dynamics is insensitive to the precise choice of microstate.

\section{Summary}

The black hole information paradox has taken us on an interesting journey, at the end of which we seem to have a remarkable change in our perspective on gravity. The first thing to note is that the paradox is not {\it trivial}. Many people had tried to ignore the paradox hoping that small corrections to Hawking's computation would lead to information being 	`delicately encoded' in the outgoing radiation. The inequality (\ref{seven}) shows that this belief is {\it false}; no small corrections to Hawking's computation can bypass the information problem. 

\begin{figure}[htbt]
\hskip .5 in\includegraphics[scale=.45]{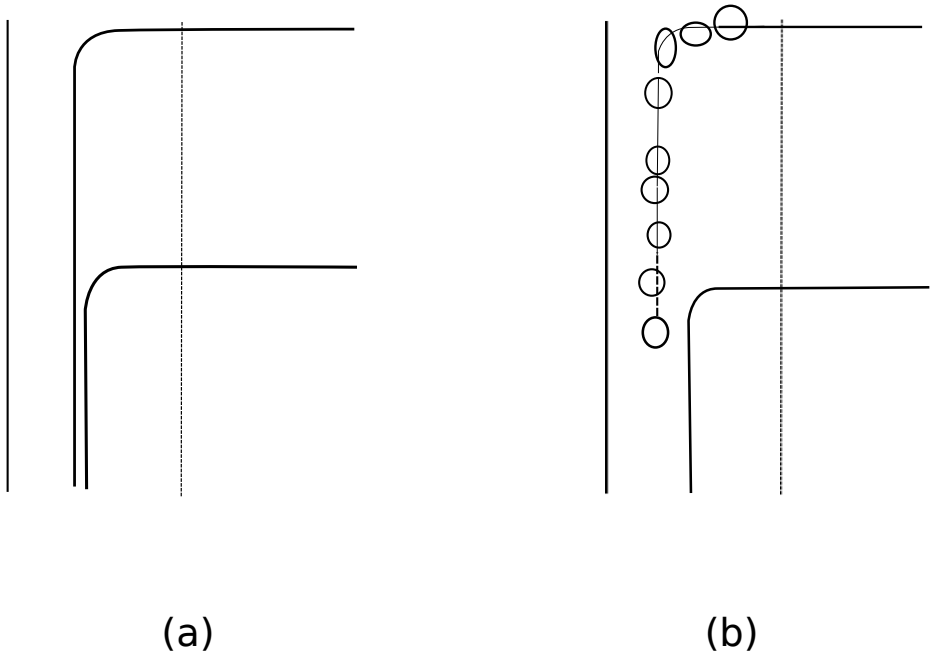}
%
%
\caption{(a) The stretching of `good slices' in the traditional black hole geometry leads to pair creation by the Hawking process and the consequent information problem. (b) If there are $Exp[S]$ fuzzball solutions, the wavefunction giving semiclassical geometry on the initial slice spreads over this vast phase space of solutions after some evolution, and we no longer get the traditional pair creation with growing entanglement.}
\label{fa13}       
\end{figure}

But if the paradox is nontrivial, then we must necessarily find some fundamental change in our basic assumptions about physics when we do solve the paradox. And this is what we find when we use string theory: the problem is removed because the entire structure of the black hole undergoes a fundamental revision. In fig.\ref{fa13}(a) we depict again the `good slicing' of the traditional black hole that was used to derive the paradox in section \ref{secc}. The slices `stretch' indefinitely inside the horizon, and each step of stretching leads to one unit of pair production. Since the curvature is low all along the slice, we can find no way to avoid the consequences of this evolution if the traditional black hole geometry is correct.

In fig.\ref{fa13}(b) we depict schematically how the situation changes in string theory. There are $Exp[S_{bek}]$ fuzzball microstates with no horizon, and the semiclassical vacuum state on the initial slice spreads over this vast phase space as we evolve the slice through the stretching process involved in the derivation of Hawking radiation. The smallness of the transition amplitude to any one fuzzball state is offset by the large number of fuzzball states; here we make explicit use of the enormous entropy that sets black holes apart from all normal objects. 

Thus the essential new physics is the spread of the wavefunction of a collapsing shell over the vast available space of gravitational solutions  with the same quantum numbers. But if this lesson is correct, then we should expect new physics in other situations where we have a large mass confined to a given region; for example in the very early Universe when matter densities are expected to be very high. Should there be an exponentially large number of gravitational solutions again with the same quantum numbers? If so, the wavefunction spread over this vast phase space can change classical dynamics by order unity, and in particular may resolve the initial singularity in the same way that we have avoided the black hole central singularity. Some initial considerations in this direction are discussed in  \cite{cmuniv}. 

Allowing information escape required us to find a way that different microstates would behave differently at their boundary, and this the fuzzball construction achieves. But we can also ask the opposite question: is there any sense in which all generic fuzzballs behave the {\it same}, and if so, can this universal behavior be represented by the traditional black hole geometry? Putting fuzzballs in the context of ideas developed in \cite{israel,eternal,raamsdonk} we found that the answer is {\it yes}. A central role in this analysis was played by the relation \cite{raamsdonk} depicted in fig.\ref{fn7}: summing over fuzzball states reproduces a spacetime that has smooth horizons at the region where individual fuzzballs end in a very messy boundary with string sources. 
An individual black hole microstate is just one fuzzball, and so has no horizon or region interior to the horizon. But for any process where we can replace the fuzzball by a {\it sum} over fuzzballs -- i.e. we can take a thermodymamic average -- we can use the traditional black hole geometry with smooth horizons. Using such a picture we postulated solutions to many puzzles in gravity. What is the meaning of Rindler and de Sitter entropies? How can we reconcile the idea of `complementarity' with normal evolution on Cauchy slices? What is the fate of an object falling through the horizon? 

To summarize, analyzing black holes  using string theory has provided a very rich set of new ideas. These ideas may help us understand many other deep questions in fundamental physics, particularly those associated to the very early Universe.

 \section*{Acknowledgements}

This work puts together ideas developed over several decades by the entire community of people working on black holes. I would like to thank all these folks. In particular, I would like to thank all my collaborators who have helped me find my way through this exciting field. 
This work was supported in part by DOE grant DE-FG02-91ER-40690.

\end{document}